\documentclass{article}
\usepackage[letterpaper,top=2cm,bottom=2cm,left=3cm,right=3cm,marginparwidth=1.75cm]{geometry}

\usepackage{braket,amsfonts}
\usepackage{graphicx}
\usepackage{amsmath}
\usepackage{amssymb}
\usepackage{amsfonts}
\usepackage[caption=false]{subfig}
\usepackage{pgfplots}
\usepackage{algorithm}
\usepackage[noend]{algpseudocode}
\usepackage{graphics}
\usepackage{color}
\usepackage{epsfig}
\usepackage{cite}
\usepackage{float}
\usepackage{wrapfig}
\usepackage{bbold}
\usepackage{appendix}
\usepackage{textcomp}
\usepackage{graphicx,color}
\usepackage{mathrsfs}
\usepackage{url}
\usepackage{color}
\usepackage{dsfont}
\usepackage{bbm}
\usepackage{verbatim}
\usepackage{tikz}
\usepackage{titlesec}
\usepackage{abstract}
\usepackage{caption}
\usepackage{titlesec}
\usepackage{bm}

\usepackage{xr}

\titleformat*{\section}{\LARGE}
\titleformat*{\subsection}{\Large}

\newcommand{\real}{{\mathbb{R}}}

\newcommand{\x}{\operatorname{x}}
\newcommand{\vctrl}{\operatorname{v}}
\newcommand{\ctrl}{\operatorname{u}}
\newcommand{\ctrlvec}{\underline{\bf u}}
\newcommand{\xvec}{\underline{\bf x}}
\newcommand{\vecx}{{\bf x}}

\newcommand{\cvec}{\underline{c}}
\newcommand{\zvec}{\underline{\bf z}}

\newcommand{\vvec}{\underline{\bf v}}
\newcommand{\xivec}{\underline{\boldsymbol{\xi}}}
\newcommand{\etavec}{\underline{\boldsymbol{\eta}}}
\newcommand{\phivec}{\underline{\boldsymbol{\phi}}}
\newcommand{\lambdavec}{\underline{\boldsymbol{\lambda}}}
\newcommand{\muvec}{\underline{\boldsymbol{\mu}}}
\newcommand{\nuvec}{\underline{\boldsymbol{\nu}}}

\makeatletter
\newcommand*{\addFileDependency}[1]{
  \typeout{(#1)}
  \@addtofilelist{#1}
  \IfFileExists{#1}{}{\typeout{No file #1.}}
}
\makeatother

\title{Hamiltonian bridge: A physics-driven generative framework for targeted pattern control}
\author{Vishaal Krishnan$^{1, \dagger}$ \quad  Sumit Sinha$^{1,2 \dagger}$ \quad  L. Mahadevan$^{1,3\ast}$
\\
\footnotesize{$^1 $ School of Engineering and Applied Sciences, Harvard University, Cambridge, MA 02138, USA}\\
\footnotesize{$^2 $ Department of Data Science, Dana-Farber Cancer Institute, Boston, MA, USA}\\
\footnotesize{$^3$  Departments of Physics, and Organismic and Evolutionary Biology, Harvard University, Cambridge, MA 02138, USA}\\
\footnotesize{$^\ast$To whom correspondence should be addressed; E-mail:  lmahadev@g.harvard.edu}\\
\footnotesize{$^\dagger$Equal Contribution}
}
\date{}

\begin{document}

\maketitle

\begin{abstract}
Patterns arise spontaneously in a range of systems spanning the sciences, and their study typically focuses on mechanisms to understand their evolution in space-time. Increasingly, there has been a transition towards controlling these patterns in various functional settings, with implications for engineering. Here, we combine our knowledge of a general class of dynamical laws for pattern formation in non-equilibrium  systems, and the power of stochastic optimal control approaches to present a framework that allows us to control patterns at multiple scales, which we dub the ``Hamiltonian bridge". We use a mapping between stochastic many-body Lagrangian physics and deterministic Eulerian pattern forming PDEs to leverage our recent  approach utilizing the Feynman-Kac-based adjoint path integral formulation for the control of interacting particles and generalize this to the active control of patterning fields. We demonstrate the applicability of our computational framework via numerical experiments on the control of phase separation with and without a conserved order parameter, self-assembly of fluid droplets, coupled reaction-diffusion equations and finally a phenomenological model for spatio-temporal tissue differentiation. We interpret our numerical experiments in terms of a theoretical understanding of how the underlying physics shapes the geometry of the pattern manifold, altering the transport paths of patterns and the nature of pattern interpolation. We finally conclude by showing how optimal control can be utilized to generate complex patterns via an iterative control protocol over pattern forming pdes which can be casted as gradient flows. All together, our study shows how we can systematically build in physical priors into a generative framework for pattern control in non-equilibrium systems across multiple length and time scales.   
\end{abstract}





\section*{Introduction}
Over the last century, the study of the spatio-temporal dynamics of patterns started to come of age via the realization that certain general mathematical frameworks transcend specific applications, an idea that was first espoused by Landau~\cite{LL:37} in the context of equilibrium phase transitions, but since generalized to other physical~\cite{MC-HG:09}, chemical~\cite{RD-RK:09}, material~\cite{RB-SA-CC:05} and biological systems~\cite{JDM:07}. Theories for pattern prediction follow from  different microscopic dynamics, e.g. Langevin dynamics or master-equation frameworks that take phenomenological coarse-grained forms of conservation laws and various types of closures guided by symmetry (and broken-symmetry).  This typically (data-poor) theory-rich approach which has had enormous success in explaining the origin of patterns in natural and artificial settings has recently begun to be supplanted and supplemented by a  data-rich (theory-poor) approach to pattern generation driven by advances in modern machine learning methods that enable the generation of diverse and realistic patterns by learning from large datasets~\cite{JSD-EW-NM-SG:15}. Recent work~\cite{GB-TB-VDB-MM:24} has attempted to provide insights into the underlying mechanisms driving pattern emergence in these models which leverage the combination of vast training sets and fast computing methods and resources. 
This naturally raises the question of how to build on this and incorporate the principles of pattern formation embodied in conservation laws, symmetry constraints, and material properties (empirical constitutive relations) as priors for generative algorithms for pattern control? And thence incorporate interpolation schemes that allow for the specification of physical constraints and dynamical laws into algorithms for pattern control?

Techniques for interpolation fall broadly into the classes that use microscopic stochastic approaches~\cite{MA-NB-EVE:23, YC-TG-MP:16} or coarse-grained   hydrodynamic approaches~\cite{JDB-YB:00}, some of which have variational foundations~\cite{RJ-DK-FO:98}. An incomplete list includes generative models based on optimal transport~\cite{LR-AK-EB:21, DA-YG-NL-ZL-STY-XG:19}, unbalanced optimal transport~\cite{LC-GP-BS-FXV:18}, normalizing flows~\cite{EGT-EVE:10,EGT-CVT:13,DR-SM:15} and diffusion models~\cite{JSD-EW-NM-SG:15,YS-JSD-DPK-AK-SE-BP:20}. The shift in perspective from the forward problem of determining the evolution of patterns given local interactions, towards the inverse problem of evoking global pattern forming functionality from local interactions needs to accommodate both the microscale processes that drive the pattern and the macroscale patterns that determine functionality, e.g. in engineering and biology, where molecular interactions control morphology that both enables and constrains function. In the context of  pattern control, this interplay is even more acute, as while the patterns may be functional at one scale, control may need to be distributed across several scales, resulting in a complex multiscale interdependence.
Here, we focus our attention on active control of long-wavelength patterns. We employ optimal control and optimal transport theories to cast the problem of active pattern control as one of interpolation between the initial and target patterns, which we call the Hamiltonian bridge. This results in a principled generative framework to interpolate between different patterned states while respecting physical constraints that take the form of dynamical laws, thereby enabling the synthesis of physically meaningful functional patterns. This formulation reveals key aspects of the geometry of pattern formation and how it is shaped by the underlying physics. Building on our previous work on controlling interacting particles~\cite{SS-VK-LM:23}, here we consider the control of fields that arise naturally in several pattern forming systems in soft and active matter, including phase separation with and without a conserved order parameter, the hydrodynamics of droplets, coupled reaction-diffusion equations, and models for cell fate determination.

\section*{Hamiltonian bridge formulation for pattern control}
We begin by considering an evolution equation that is robust enough to encompass a class pattern forming systems (see SI section S4 for derivation of deterministic PDEs from stochastic many-body lagrangian physics which also forms the algorithmic basis for solving the optimal control experiments)
\begin{align} \label{eq:pattern_formation_general_form}
    \frac{\partial \phi}{\partial t}(t, {\bf z}) + \nabla \cdot \left( m(\phi(t, {\bf z})) ( \mathbf{v}(t, {\bf z}) + \underbrace{\mathbf{u}(t, {\bf z})}_{\rm Flux~control} ) \right) = R(t, {\bf z}) + \underbrace{S(t, {\bf z})}_{\rm Reaction~control} 
\end{align}
where the scalar field $\phi(t, \mathbf{z})$ represents the system state, i.e., pattern evaluated at time~$t$ at a point~$\mathbf{z} \in \Omega$ in an (unbounded) planar domain. Here,  $m(\phi(t, {\bf z}))$ is the (potentially nonlinear) mobility, $\mathbf{v}(t, {\bf z})$ is the passive (known) flux associated with variations in the passive chemical potential, and $ \mathbf{u}(t, {\bf z})$ is the active conserved (unknown) control flux and $R(t, {\bf z})$ is the passive non-conserved (known) reaction field, and $ S(t, {\bf z})$ is the active (unknown) control reaction field, all of which are functions of  time~$t$ and space ~$\mathbf{z} \in \Omega$ defining the domain. In studies of pattern formation~\cite{cross1993pattern}, there are no active fluxes or reactions, and instead one aims to describe the phase diagrams for various types of patterns as a function of the problem parameters (e.g. the free energy, advective flow etc.). Complementing these efforts, in studies of active systems \cite{MCM-JFJ-SR-TBL-JP-MR-RAS:13}, the active fluxes and forces are known and subject to additional constraints, e.g. reciprocity, odd-ness etc. but the focus still remains on pattern prediction to define and describe the different qualitative phases that arise.  

Here, we deviate from this perspective of pattern prediction to that of pattern control and ask how we might deploy the control vector flux $\mathbf{u}(t, {\bf z})$ and/or the reaction field $S(t, {\bf z})$ to pattern $\phi(t, {\bf z})$ over a time horizon $[0,T]$ in a domain~$\Omega$ to steer the pattern towards a target pattern~$\phi^*({\bf z})$ while minimizing a measure of the steering costs. Of the various choices of the instantaneous steering costs, we choose a minimally effective one given by 
\begin{align*}
C_{\gamma_{\ctrl}, \gamma_{\vctrl}}(t, \phi, \mathbf{u}, S) = \frac{\gamma_{\ctrl}}{2} \int_\Omega m(\phi(t, \mathbf{z})) \left\| \mathbf{u}(t, {\bf z}) \right\|^2 d^2 \mathbf{z} + \frac{\gamma_{\vctrl}}{2} \int_\Omega  \left( S(t, {\bf z}) \right)^2  d^2 \mathbf{z}    
\end{align*}
where the first term is a time-extensive cost that is proportional to the flux, while the second time-extensive cost corresponds to the reaction term; $\gamma_{\ctrl}, \gamma_{\vctrl}$ are the weight parameters. The choice of a quadratic form for the cost is common in many applications, and leads to an easily implementable fast algorithm while not being not particularly restrictive. Then, the active pattern control problem  can then be formulated as~follows
\begin{align} \label{eq:opt_ctrl_pattern_formation_terminal_constraints}
    \min_{\mathbf{u}, S} ~  \int_0^T C_{\gamma_{\ctrl}, \gamma_{\vctrl}}(t, \phi, \mathbf{u}, S) dt \quad \text{s.t.}~ \begin{cases} \frac{\partial \phi}{\partial t} (t, {\bf z}) + \nabla \cdot \left( m(\phi(t,{\bf z})) ( \mathbf{v}(t, {\bf z}) + \mathbf{u}(t, {\bf z}) ) \right) = R(t, {\bf z}) + S(t, {\bf z}) \\
    \phi(0, {\bf z}) = \phi_0({\bf z}), ~\phi(T,{\bf z}) = \phi^*({\bf z})
    \end{cases}
\end{align}
For ease of notation, we drop the argument~$(t, {\bf z})$ from the PDEs as is clear from context. This framing is the well-known optimal control problem~\cite{FT:10} for infinite-dimensional systems, now reworded in the context of fields. The optimal strategy for pattern control is given by the solution to Problem~\eqref{eq:opt_ctrl_pattern_formation_terminal_constraints}, and described by a Hamiltonian system (see SI Section S2 for the derivation and~\cite{FT:10} for a detailed exposition)
\begin{align} \label{eq:hamilton_bridge}
    \frac{\partial \phi}{\partial t} = \left \lbrace \phi, H \right \rbrace, \qquad
    \frac{\partial \Lambda}{\partial t} = \left \lbrace \Lambda, H \right \rbrace
\end{align}
where $\Lambda$ is the co-state corresponding to~$\phi$, $\lbrace \cdot , \cdot \rbrace$ is the usual Poisson bracket with $\left \lbrace \phi, H \right \rbrace = \frac{\delta H}{\delta \lambda}(\phi, \Lambda)$ and $\left \lbrace \Lambda, H \right \rbrace = -\frac{\delta H}{\delta \varphi}(\phi, \Lambda)$ and the control Hamiltonian~$H$ is given by
\begin{align} \label{eq:hamiltonian}
    H \left(\varphi, \lambda \right) = \int_\Omega \left[ m(\varphi) \nabla \lambda \cdot \mathbf{v} + \lambda R - \frac{m(\varphi)}{2\gamma_{\ctrl}} \left\| \nabla \lambda \right\|^2 - \frac{\lambda^2}{2\gamma_{\vctrl}} \right] d^2 \mathbf{z}.
\end{align}
To clarify the notation adopted in the paper, we note that $(\phi(t, \mathbf{z}), \Lambda(t, \mathbf{z}))$ denotes a time-dependent state--co-state trajectory whereas $(\varphi(\mathbf{z}), \lambda(\mathbf{z}))$ denotes any state--co-state pair (in the present context, the state and co-state spaces are function spaces on the domain~$\Omega$, and therefore~$\varphi$ and~$\lambda$ are functions on~$\Omega$). 
It is important to emphasize that even though the dynamical equation~\eqref{eq:pattern_formation_general_form} is not derivable from a Hamiltonian, the augmented control framework still fits into the Hamiltonian framework and can be reformulated as
\begin{align} \label{eq:Hamiltonian_bridge}
    \max_{\Lambda_0} ~\int_\Omega \Lambda_0 \phi_0 ~d^2 \mathbf{z} - \int_\Omega \Lambda_T \phi^* ~d^2 \mathbf{z} ,
    \quad \text{s.t.} \quad 
    \frac{\partial \phi}{\partial t} = \left \lbrace \phi, H \right \rbrace, ~
    \frac{\partial \Lambda}{\partial t} = \left \lbrace \Lambda, H \right \rbrace
\end{align}
where $\phi(0, {\bf z}) = \phi_0({\bf z})$, $\Lambda(0, {\bf z}) = \Lambda_0({\bf z})$ and $\Lambda(T, {\bf z}) = \Lambda_T({\bf z})$. We refer to the optimal interpolant between~$\phi_0$ and~$\phi^*$ generated by the Hamiltonian system~\eqref{eq:Hamiltonian_bridge} as the Hamiltonian bridge. 
\begin{figure}[!h]
    \centering
    \includegraphics[width=0.8\linewidth]{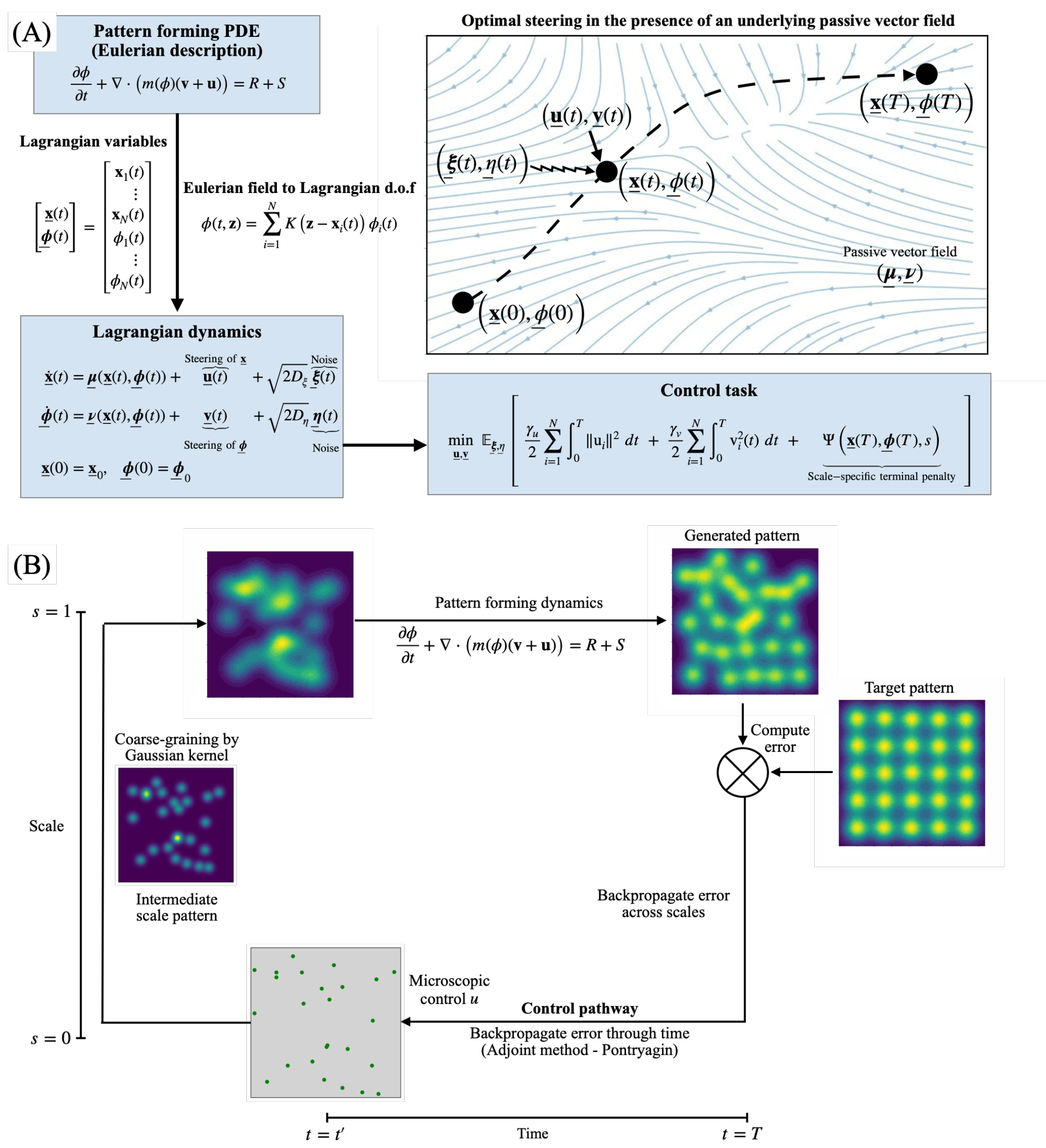}
    \caption{\textbf{Algorithmic framework for targeted pattern control.}\textbf{(A)} Physics-driven framework for pattern generation. Top-left panel shows the general form of conserved and non-conserved PDEs considered in the present study. Here, $\varphi$ can be a scalar or vector valued field, $\mathbf{v}$ is a known vector field modeling the intrinsic pattern forming capacity and $\mathbf{u}$ is the control flux field, $R$ is a given reaction field, and $S$ is the control reaction field as described in~\eqref{eq:pattern_formation_general_form}. The pattern control objective is formulated as the steering of the pattern generated by the forward dynamics towards a target pattern as described in~\eqref{eq:opt_ctrl_pattern_formation_terminal_constraints}. In the Eulerian-Lagrangian control formulation, the Eulerian field $\phi(t, {\bf z})$ is first discretized via the Smoothed Particle Hydrodynamics (SPH) method (see SI section S4) to Lagrangian dofs, ${\bf x}_i$ and $\phi_i$, which when deployed on the Eulerian PDE, leads to a set of coupled non-linear equations for the Lagrangian dofs (see SI section S4). The Lagrangian dofs are then steered via ${\bf u}$ and ${\bf v}$ to obtain the target field via the optimal control objective. The Eulerian-Lagrangian optimal control formulation can be viewed as navigation on a complex landscape as shown on the right. \textbf{(B)} Schematic of the Lagrangian-Eulerian optimal control method for pattern design. At time $t'$, consider a Lagrangian set of coordinates as shown in bottom-left corner which gives a Eulerian pattern via smoothing kernels. Utilizing the Lagrangian degrees of freedom (dofs), the Eulerian fields evolve in time via the forward dynamical equations. At the terminal time, the error between the target pattern and evolved pattern is computed and backpropagated (via deterministic or stochastic adjoint method) both across scales and in time, to obtain a control at time $t$, which allows the Lagrangian coordinates (and the Eulerian pattern) to evolve in time (for details see SI Section S4).}
    \label{fig:intro}
\end{figure}

%

\section*{Eulerian-Lagrangian optimal control framework} 
With the general framework in place, we now briefly describe the numerical method for implementing the Hamiltonian bridge formulation~\eqref{eq:Hamiltonian_bridge} for optimally steering pattern forming systems (for details see SI section S4). Since the control of patterns often involves moving boundaries, we adopt a Lagrangian perspective instead of an Eulerian description. We switch between the Eulerian representation given by (1) and the Lagrangian representation using smoothed-particle-hydrodynamics~\cite{JJM:12}, an approach to convert any field-theoretic formulation into a particle-based meshless formulation that is amenable to parallelization and speed-up. This allows us to discretize the forward (passive) Eulerian PDE and create a Lagrangian framework for the dynamics of particles and write the patterning field as $\phi(t, \mathbf{z}) = \sum_{i=1}^N \kappa({\bf z}-{\bf x}_i(t)) \phi_i(t)$, where $\kappa({\bf z}-{\bf x}_i(t)) \sim \exp(-\frac{||{\bf z}-{\bf x}_i(t)||^2}{\sigma(s)^2})$ is an unnormalized Gaussian kernel with $\sigma(s)^2$ being the size of the ``smoothed-particle" with $s \in [0,1]$ characterizing the coarse-graining scale (with $s=0$ being maximally fine-grained and $s=1$ being maximally coarse-grained),  ${\bf z}$ the discretized Eulerian position in~$\Omega$, $\xvec(t) = ({\bf x}_1(t), \ldots, {\bf x}_N(t))$ the dynamic Lagrangian positional degrees of freedom (${\bf x}_i(t) \in \Omega$ for all $i \in \{1, \ldots, N \}, t \in \real_{+}$),  and $\phivec(t) = (\phi_1(t), \ldots, \phi_N(t))$ is the discretized Lagrangian pattern field (as shown in Figure~\ref{fig:intro}A). Importantly, the particle-based representation of the forward problem for pattern evolution allows us to utilize our recently developed Feynmac-Kac path integral adjoint method for many-body control~\cite{SS-VK-LM:23} on the Lagrangian dofs $(\xvec(t), \phivec(t))$ to solve for the control field to steer the pattern dynamics to the desired end state (Figure~\ref{fig:intro}B for a schematic of the implementation of pattern control). Our Eulerian-Lagrangian control formalism naturally extends to situations where the pattern can be steered in a scale dependent (i.e $s$ can be a continuous variable from $0$ to $1$) fashion which can accommodate for concepts like renormalization \cite{JC-SR:23,TM-MO-GB-SM:22}.

\section*{Targeted steering of pattern forming systems}
Motivated by the problems associated with phase separation kinetics, droplet movements on substrates and reaction-diffusion systems that arise in a number of areas in the sciences and engineering, e.g. physical chemistry, materials science, active matter, biophysics etc. we focus here on a few paradigmatic cases: (a) the Allen-Cahn equation describing the dynamics of systems with a non-conserved order parameter, i.e. Model A ~\cite{PCH-BIH:77,RB-SA-CC:05}, (b) the Cahn-Hilliard equation describing the dynamics of systems with a conserved order parameter, i.e. Model B ~\cite{PCH-BIH:77,RB-SA-CC:05}, (c) the thin liquid film dynamics associated with drops on a substrate, modeled by a variant of the Model B equations with a nonlinear mobility \cite{TW:20,EBD-VD-SHD:74}, (d) a minimal set of coupled reaction-diffusion equations underlying the classical Turing instability~\cite{JDM:07}, and finally (e) a phenomenological model for cell-fate dynamics in tissue differentiation that couples the (external) positional degrees of freedom with an (internal) fate.

A common theme among all the above examples that we exploit is the ability to interpret the forward (uncontrolled) problem associated with pattern formation as a gradient flow. Then, we can write (1) in the absence of control fields, i.e. with $ \mathbf{u}(t, {\bf z})=0$ and $ S(t, {\bf z})=0$,  succinctly as
\begin{align}
    \frac{\partial \phi}{\partial t} = D_\xi \Delta \phi + \beta_1 \nabla \cdot \left( m(\phi) \nabla \left( \frac{\delta E_1}{\delta \varphi}  \bigg|_{\phi} \right) \right) - \beta_2 \frac{\delta E_2}{\delta \varphi}\bigg|_{\phi},
    \label{drops_eqn}
\end{align}
$D_\xi = k_B T/ \gamma_{\rm u}$ being the diffusion constant, $\Delta$ is the Laplacian. Comparing (1) and (6), we see that this implies that the flux $\mathbf{v} = - \beta_1 \nabla \left( \frac{\delta E_1}{\delta \varphi} \right)$, and the reaction $R = - \beta_2 \frac{\delta E_2}{\delta \varphi}$, where $\beta_1, \beta_2$ are constants and $E_1, E_2$ are energy functionals of the form
\begin{align} \label{eq:free_energy1}
    E_k(\varphi) = \int_{\Omega} \left[ U_k(\varphi(\mathbf{z})) + \frac{\epsilon_k}{2} \left\| \nabla \varphi(\mathbf{z}) \right\|^2 \right] d^2 \mathbf{z}, \qquad k=1,2 
\end{align}
where $U_k(\varphi(\mathbf{z}))$ is the potential energy density, and $\frac{\epsilon_k}{2} \left\| \nabla \varphi(\mathbf{z}) \right\|^2$ is the term which penalizes gradients  with $\epsilon_k$ being a constant (akin to surface-tension), with the variational derivative $ \frac{\delta E_k}{\delta \varphi} = U_k' - \epsilon_k \Delta \varphi$ and  $U_k'=\frac{dU_k}{d\varphi}$.  While it might initially appear unduly restrictive to consider forward physics governed only by gradient flows, we shall explain later on that the framework of optimal control allows for encoding pattern forming functionality as gradient flows across a very wide range of problems.
%

In the limit  $D_\xi = 0$, with $\beta_1 = 0 , \beta_2 > 0$, we note that the dynamics~\eqref{drops_eqn} does not conserve the order-parameter, i.e. $\int_\Omega \varphi d^2 \mathbf{z}$ evolves under the dynamics~\eqref{drops_eqn}. With $U_2(\varphi) = \frac{1}{2} \left( 1- \varphi^2 \right)^2$ as the free energy density of phase separation, we obtain the Model-A class of equations \cite{PCH-BIH:77}, e.g. the Allen-Cahn equation for phase separating mixtures. When  $\beta_1 > 0 , \beta_2 = 0$, we note that the dynamics~\eqref{drops_eqn} conserves the order-parameter, i.e. $\int_\Omega \varphi d^2 \mathbf{z}$ is a constant; when $U_1(\varphi) = \frac{1}{2}  \left( 1- \varphi^2 \right)^2$ is the free energy density of phase separation,  we obtain  the well-known Model-B class of equations for phase separating mixtures \cite{PCH-BIH:77}.  When the mobility $m(\varphi) = m$ is a constant, this corresponds to the Cahn-Hilliard equation, and when the mobility $m(\varphi)\propto \varphi^3$, we recover a variant of the thin-film hydrodynamic equation \cite{EBD-VD-SHD:74}, wherein $\varphi$ describes the height of a thin film of liquid. It is worth noting that a number of active matter systems lead to very similar forms of phase-separation 
kinetics~\cite{tailleur2022active}, and so are also amenable to our framework, although we limit ourselves here to the commonest of these cases.

\subsection*{Controlling phase separation}
We start by considering the case of controlling phase separation for the case of the dynamics associated with either a non-conserved order parameter (Model A: Allen-Cahn), or a conserved order parameter (Model B: Cahn-Hilliard). For Model A dynamics (Allen-Cahn), the (uncontrolled) phase field $\phi(t,\mathbf{z})$ satisfies the equation $\phi_t = \phi - \phi^3+ \nabla^2 \phi$ (and corresponds to the energy functional $E(\varphi)=\int_{\Omega} \left[ (1-\varphi^2)^2 + \frac{\epsilon}{2} \left\| \nabla \varphi(\mathbf{z}) \right\|^2 \right] d^2 \mathbf{z}$, with $\mathbf{v} = 0$ and $R = - \frac{\delta E}{\delta \varphi}$ in (1).  For Model B dynamics (Cahn-Hilliard),  the (uncontrolled) phase field $\phi(t,\mathbf{z})$ satisfies the equation $\phi_t = \nabla^2 (-\phi + \phi^3 - \nabla^2 \phi)$ (corresponding to the velocity field $\mathbf{v}=-\nabla \frac{\delta E}{\delta \varphi}$ and the reaction field $R = 0$ in Eq. 3).

\begin{figure}[!h]
    \centering
    \includegraphics[width=1.0\linewidth]{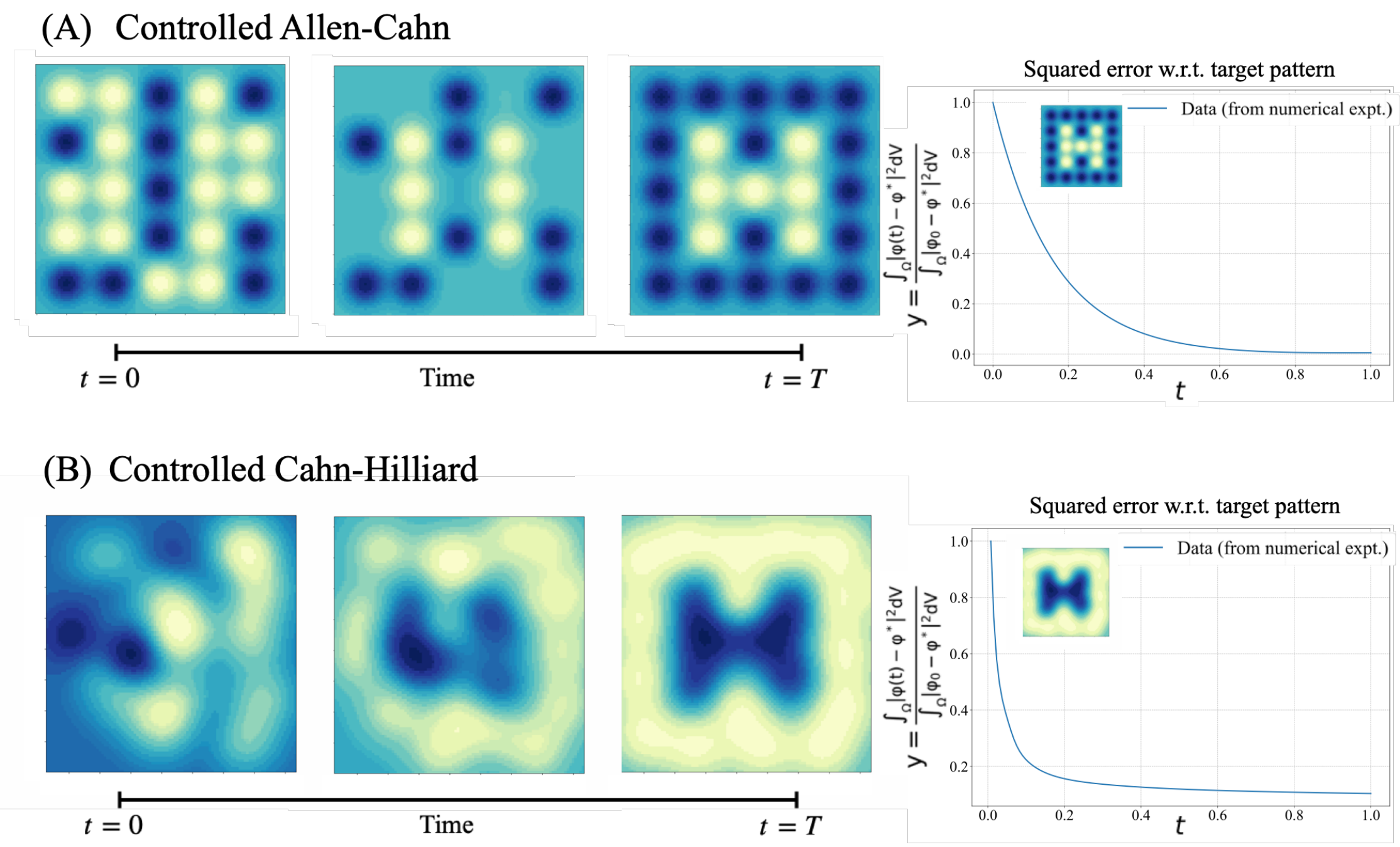}
    \caption{{\bf Targeted phase separation.} {\bf (A)} Controlling the phase separation of two species (blue and white colors in a square lattice chequered board pattern) governed by the Allen-Cahn Equation (see \ref{drops_eqn} and surrounding text). The uncontrolled Allen-Cahn equation is given by $\frac{\partial \phi}{\partial t} =R$, where $R$ is the reaction field given by $R=- \frac{\delta E }{\delta \varphi}$, where $E(\varphi)=\int_{\Omega} \left[ U(\varphi(\mathbf{z})) + \frac{\epsilon}{2} \left\| \nabla \varphi(\mathbf{z}) \right\|^2 \right] d^2 \mathbf{z}$ and  $U(\varphi)=(1-\varphi^2)^2$, $\epsilon=10^{-3}$. The Eulerian pattern error and the target pattern `H' (inset) is shown on the rightmost panel. 
    {\bf (B)} Controlling the phase separation of two species (blue and white colors from a randomly distributed initial condition) governed by the Cahn-Hilliard equation (see \ref{drops_eqn} and surrounding text) .  The uncontrolled Cahn-Hilliard equation is given by $\frac{\partial \phi}{\partial t} + \nabla \cdot \left( m(\phi)  \mathbf{v} \right)=0$, where $\phi$ is the phase, $m(\varphi)=m$ is the mobility, and $\mathbf{v}$ is the known vector field given by $\mathbf{v}=-\nabla \frac{\delta E }{\delta \varphi}$, where $E(\varphi)=\int_{\Omega} \left[ U(\varphi(\mathbf{z})) + \frac{\epsilon}{2} \left\| \nabla \varphi(\mathbf{z}) \right\|^2 \right] d^2 \mathbf{z}$ and  $U(\varphi)=(1-\varphi^2)^2$, $\epsilon=10^{-3}$.  The Eulerian pattern error and the target pattern `H' (inset) is shown on the rightmost panel.}
    \label{fig:phasesep}
\end{figure}

In either case, we assume that the initial condition is a random distribution given by $ \phi_0$ (see leftmost panels in Figure~\ref{fig:phasesep}A and Figure~\ref{fig:phasesep}B) and the final target distribution, $\phi^*$, is in the form of the letter `H' (see top left in the bottom panels of Figure~\ref{fig:phasesep}A and Figure~\ref{fig:phasesep}B). We see that as the (flux/reaction) control fields  pushes the pattern towards the target, the error decreases approximately exponentially (power law) in time for Allen-Cahn (Cahn-Hilliard)(see the inset  SI Figure S1, and Movie1  and Movie2  for animations of the controlled Allen-Cahn and Cahn-Hilliard systems respectively).  
To interpret these results, we use the theoretical formalism of the Hamiltonian bridge Eqns.(2-4) to describe the geometry of optimal paths/interpolants connecting the initial pattern $\phi_0({\mathbf z})$ to the final state $\phi^*({\mathbf z})$.

%

\vspace{0.1in}
\noindent \textbf{(A) Reaction field determines the nature of the interpolant for control of non-conserved order parameter dynamics (Allen-Cahn).} In the absence of any intrinsic pattern forming capability in a system with a non-conserved order parameter, $R=0, \mathbf{v} = 0$, so that the target pattern can be fully steered via a non-conserved dynamics associated with a control reaction field $S$. Then the optimal control reaction field is given by $S = - \frac{1}{\gamma_{\rm v}} \Lambda$, where $\frac{\partial \Lambda}{\partial t}(t, \mathbf{z}) = 0$, i.e. $\Lambda(t, \mathbf{z}) = \Lambda_0(\mathbf{z})$ (see SI Section S3.1 for details). From the terminal constraints in~\eqref{eq:opt_ctrl_pattern_formation_terminal_constraints}, we have $\phi(0, \mathbf{z}) = \phi_0(\mathbf{z})$ and $\phi(T, \mathbf{z}) = \phi^*(\mathbf{z})$, so that  $\Lambda(t, \mathbf{z}) = \Lambda_0(\mathbf{z}) = \phi^*(\mathbf{z}) - \phi_0(\mathbf{z})$, resulting in the optimal pattern evolution corresponding to a linear interpolation (for the case of nominal Hamiltonian $H_0 \left(\varphi, \lambda \right) = - \int_\Omega  \frac{\lambda^2}{2\gamma_{\vctrl}} d^2 \mathbf{z}$) between $\phi_0$ and $\phi^*$, i.e., $\phi(t, \mathbf{z}) = \phi_0(t, \mathbf{z}) + \frac{t}{T} (\phi^*(t, \mathbf{z}) - \phi_0(t, \mathbf{z}))$ for $t \in [0,T]$. 

\begin{figure}[!h]
    \centering
    \includegraphics[width=0.98\linewidth]{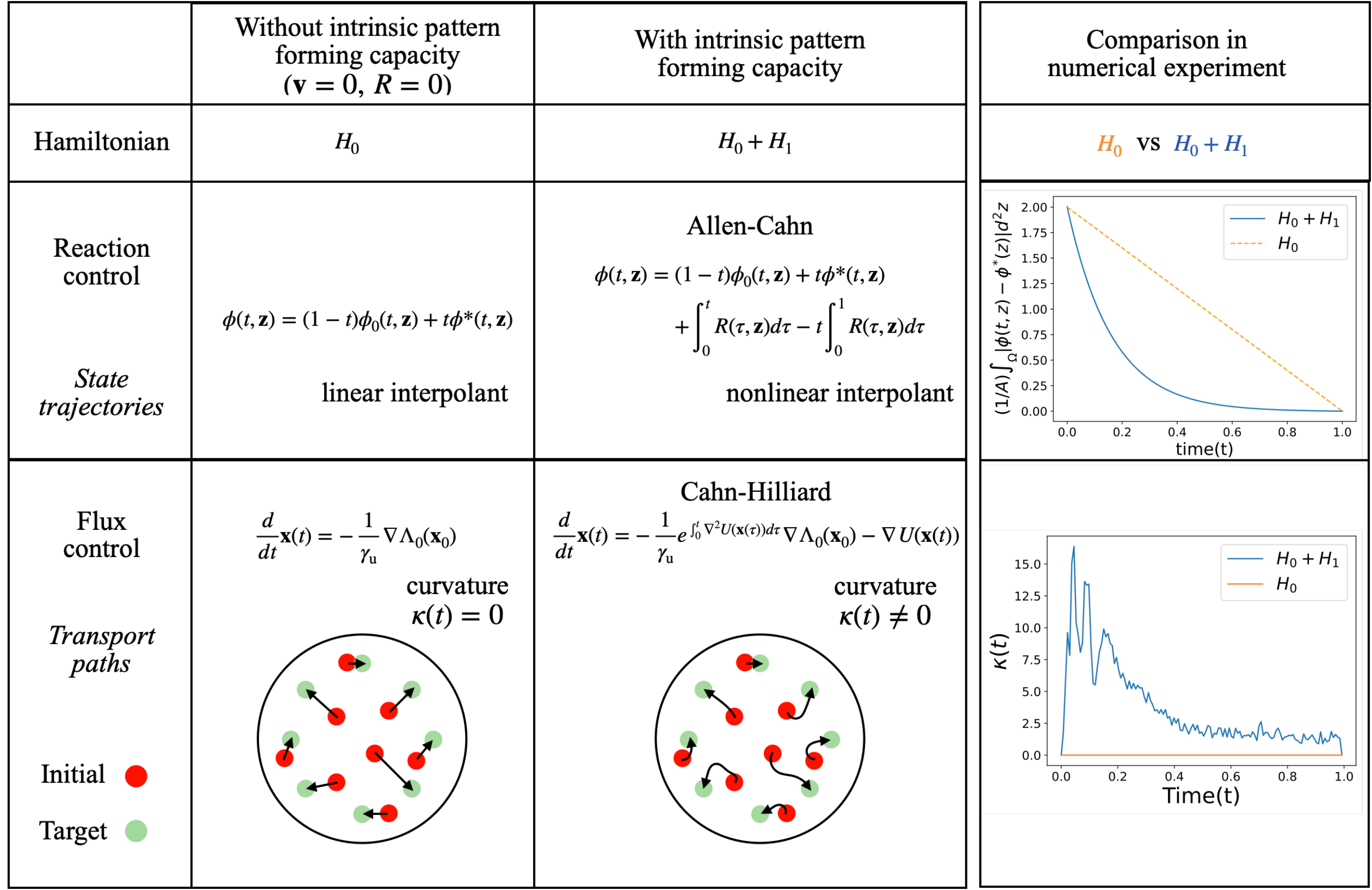}
    \caption{\textbf{Geometry of transport paths for pattern control.} Pattern forming capacity implied by the physics changes the geometry of the transport paths underlying optimal pattern evolution (see (8)-(9) and surrounding text). The row corresponding to reaction control shows the optimal interpolant in pattern space for the case of a non-conserved order parameter, corresponding to the nominal Hamiltonian $H_0 \left(\varphi, \lambda \right) = - \int_\Omega  \frac{\lambda^2}{2\gamma_{\vctrl}} d^2 \mathbf{z}$, and with intrinsic pattern forming capacity encoded in the Hamiltonian
    $H_1 \left(\varphi, \lambda \right) = \int_\Omega \lambda R d^2 \mathbf{z}$.
    The row corresponding to flux control shows the ordinary differential equation
    underlying the transport paths for the case of a conserved order parameter, corresponding to the nominal Hamiltonian $H_0 \left(\varphi, \lambda \right) =  - \int_\Omega \frac{1}{2\gamma_{\rm u}} \varphi \left\| \nabla \lambda \right\|^2 ~d^2 \mathbf{z}$, where $\varphi$ is the time-dependent pattern function and  $\lambda$ is the corresponding conjugate momentum density. The top-left panel corresponds to the zero passive reaction field which leads to a linear interpolation of the patterns. The top-middle panel corresponds to a non-linear interpolation of the pattern when the reaction field is present (control of Allen-Cahn). The bottom-middle panel corresponds to the addition of a Hamiltonian for the intrinsic pattern forming capacity $H_1\left(\varphi, \lambda \right) = - \int_\Omega m(\varphi) \nabla \lambda \cdot \nabla U ~d^2 \mathbf{z}$ (control of Cahn-Hilliard). The rightmost column corresponds to the comparison in numerical experiments. In the rightmost column, we plot the mean target error between the optimal interpolant and target pattern as a function of time for the reaction control case (top-right panel) corresponding to the Allen-Cahn dynamics, and the curvature of the transport paths as a function of time for flux control corresponding to the Cahn-Hilliard dynamics. }
    \label{fig:transport_geometry}
\end{figure}

When the passive pattern-forming capacity is present, i.e. the reaction field $R \neq 0$, we obtain a modified path (with $H_1 \left(\varphi, \lambda \right) = \int_\Omega \lambda R d^2 \mathbf{z}$) for the controlled evolution of the pattern (see SI Section S3.1 for details)
\begin{align} \label{eq:pure_non-conservative_ctrl_exp_solution}
    \phi(t, \mathbf{z}) = \phi_0(t, \mathbf{z}) + \frac{t}{T} \left( \phi^*(t, \mathbf{z}) - \phi_0(t, \mathbf{z}) \right) + \int_0^t R(\tau,\mathbf{z}) d\tau - \frac{t}{T} \int_0^T R(\tau,\mathbf{z}) d\tau. 
\end{align}
We can visualize this non-linear interpolation between $\phi_0$ and $\phi^*$ in the upper-right panel of Figure \ref{fig:transport_geometry} where the state trajectories corresponding to $R = 0$ and $R\neq 0$, i.e, Allen-Cahn, are plotted, and note that our interpolant weights the passive physics and active control to achieve the final state with a minimum control cost. \\

\noindent \textbf{(B) Effective potential determines path of pattern evolution in the case of conserved order parameter dynamics   (Cahn-Hilliard).} \label{sec:effec_pot} 
In the absence of any intrinsic pattern forming capability in a system with a  conserved order parameter, i.e. $R=0, \mathbf{v} = 0$, and $\int_\Omega \phi_0 d^2 \mathbf{z} = \int_\Omega \phi^* d^2 \mathbf{z}$, the target pattern can be fully steered via a conserved dynamics associated with a control flux ~$\mathbf{u}$.
In the context of the Hamiltonian bridge formalism, we can interpret this in terms of a nominal Hamiltonian $H_0$ and mobility $m(\varphi) = \varphi$ (which corresponds to the dynamics for the case of a conserved order-parameter case for any control vector field~$\mathbf{u}$), while the pattern-forming capacity via passive physics corresponds to an energetic contribution to the Hamiltonian $H_1$, with $H=H_0+H_1$.

To investigate how this changes the geometry of pattern evolution, we note that the nominal control Hamiltonian (in the absence of pattern-forming capacity) for a conserved order-parameter pattern evolution (see Eq. (3))  is given by $H_0 \left(\varphi, \lambda \right) =  - \frac{1}{2\gamma_{\ctrl}} \int_\Omega \varphi \left\| \nabla \lambda \right\|^2 d^2 \mathbf{z}$.
The adjoint/co-state equation then follows from~\eqref{eq:hamilton_bridge} as  $\frac{\partial \Lambda}{\partial t} = - \frac{\delta H_0}{\delta \varphi} (\phi, \Lambda)$  and simplifies to $\frac{\partial \Lambda}{\partial t} - \frac{1}{2\gamma_{\ctrl}} \left\| \nabla \Lambda \right\|^2 = 0$. This last equation for the co-state is identical to the equation for the fluid potential in the well-known Benamou-Brenier formulation of optimal transport~\cite{JDB-YB:00} that uses a hydrodynamic framework to transform normalized probability distributions via flows. Taking the gradient of the equation yields  a quasilinear equation for the gradient of the co-state ~$\frac{\partial \nabla \Lambda}{\partial t} - \frac{1}{\gamma_{\ctrl}} \nabla^2 \Lambda \nabla \Lambda = 0$, with characteristics given by the equation 
\begin{align} \label{eq:transport_paths_nominal}
    \frac{d}{dt}\mathbf{x}^{\rm nom}(t) = - \frac{1}{\gamma_{\ctrl}} \nabla \Lambda^{\rm nom}_0(\mathbf{x}^{\rm nom}_0), ~~\mathbf{x}^{\rm nom}(0) = \mathbf{x}^{\rm nom}_0,
\end{align}
along which ~$\frac{D}{Dt} \nabla \Lambda(t, \mathbf{x}(t)) = 0$, so that $\nabla \Lambda(t, \mathbf{x}(t)) = \nabla \Lambda(0, \mathbf{x}_0) = \nabla \Lambda_0(\mathbf{x}_0)$. Solving the characteristic equation allows us to infer that optimal pattern evolution occurs along straight lines (geodesics) given by $\mathbf{x}(t) = \mathbf{x}_0 - \frac{t}{\gamma_{\ctrl}} \nabla \Lambda_0(\mathbf{x}_0)$, where~$\Lambda_0$ can be computed from the terminal constraints~$\phi_0$ and~$\phi^*$ (see SI Section S3.2 for details of the derivation).  

In the presence of an intrinsic pattern-forming capacity associated with the Hamiltonian $H_1\left(\varphi, \lambda \right) = \int_\Omega \left( - \varphi \nabla \lambda \cdot \nabla U \right)d^2 \mathbf{z}$ (where for ease of notation we let the potential $U(\mathbf{z}) = \left. \frac{\delta E(\mathbf{z})}{\delta \varphi}\right|_{\varphi}$ for the energy~$E$ in~\eqref{eq:free_energy1}), the transport paths corresponding to the optimal pattern evolution under the total Hamiltonian ~$H = H_0 + H_1$ (see SI Section S3.2 for details) are given by
\begin{align} \label{eq:transport_paths_under_physics}
     \frac{d}{dt}\mathbf{x}(t) = - \nabla U(\mathbf{x}(t)) - \frac{1}{\gamma_{\ctrl}} e^{\int_0^t \nabla^2 U(\mathbf{x}(\tau)) d\tau} \nabla \Lambda_0(\mathbf{x}_0),~~\mathbf{x}^{\rm nom}(0) = \mathbf{x}^{\rm nom}_0,
\end{align}
where as before $\Lambda_0$ is specified by the terminal constraints~$\phi_0$ and~$\phi^*$ on the pattern evolution. Comparing~\eqref{eq:transport_paths_nominal} and~\eqref{eq:transport_paths_under_physics}, we see that the straight transport paths (geodesics) associated with steering an initial pattern in the absence of physics via control are reshaped in the presence of physics, embodied in a potential, with the  curvature of the transport paths determined by the Hessian of the potential~$V$~\cite{MS-AL-VB:24}. 
To see this, we note that differentiating  ~\eqref{eq:transport_paths_nominal} leads to $\frac{d^2}{dt^2} \mathbf{x}^{\rm nom}(t) = 0$, while differentiating ~\eqref{eq:transport_paths_under_physics} yields ~$\frac{d^2}{dt^2} \mathbf{x}(t) = \nabla^2 U(\mathbf{x}(t)) \nabla U(\mathbf{x}(t)) = - \nabla \left( - \frac{1}{2} \left\| \nabla U \right\|^2 \right)$. In other words, the emergent optimal path of pattern evolution in the presence of a force field is controlled by an effective potential~$V_{\rm eff}(\mathbf{z}) = - \frac{1}{2} \left\| \nabla U (\mathbf{z})\right\|^2$, identical to that seen in the optimal control of interactive active particles on complex landscapes~\cite{SS-VK-LM:23}, but now is shown to be valid for fields as well. In Figure \ref{fig:transport_geometry} (bottom-right panel), we show the effect of passive physics, i.e. with ${\bf v}\neq 0$, the geometry of transport paths is curved, with $\kappa(t)\neq 0$. \\

More generally, Model B dynamics with nonlinear forms of the mobility are known to arise in a range of systems such as dewetting transitions in droplets, motility-induced phase separation etc. and our approach is easily generalizable to those systems as well. As an example, (see SI section~S5 for details), the dewetting of a fluid on a substrate corresponds to Model B dynamics with the mobility   given by $m(\varphi)=\varphi^3$, and the dewetting velocity field is given by $\mathbf{v}=-\nabla \left(\frac{\delta E }{\delta \varphi} \right)$, where $E(\varphi)=\int_{\Omega} \left[ U(\varphi(\mathbf{z})) + \frac{\epsilon}{2} \left\| \nabla \varphi(\mathbf{z}) \right\|^2 \right] d^2 \mathbf{z}$. In this case, the transport paths corresponding to the nominal Hamiltonian will themselves  have non-zero curvature owing to the nonlinear mobility, and a geometric interpretation of the control trajectory requires us to go beyond the simple considerations here, although a numerical approach is implementable (see results shown in SI Section S5,  Figure~S3 and Movie3). \\

\subsection*{Controlling morphogenesis}
We now turn to two examples inspired by reaction-diffusion systems in chemistry and engineering\cite{DS-RS:19}, and morphogenetic systems in developmental biology  \cite{SDT-MV:22, AK-JB:23,AP-JB:23} and engineered mimics thereof. Indeed, recent experimental work shows that it is possible to control patterns in reaction-diffusion experiments using empirical methods~\cite{DS-RS:19}, and similarly in developmental systems~\cite{JR:18}, raising the question of how to do so in a principled manner. This necessitates increasing the complexity by steering patterns using both conserved (flux-like) and non-conserved (reaction-like) controls. We consider two examples: (a) a system of morphogens encoded via color e.g. R, G, B  governed by a reaction-diffusion where the reaction term has quadratic non-linearity, (b) a phenomenological model for determining cell fate control that links internal cell states with that of their external positions, i.e. their neighborhood.\\   

\begin{figure}[!h]
    \centering
    \includegraphics[width=1.0\linewidth]{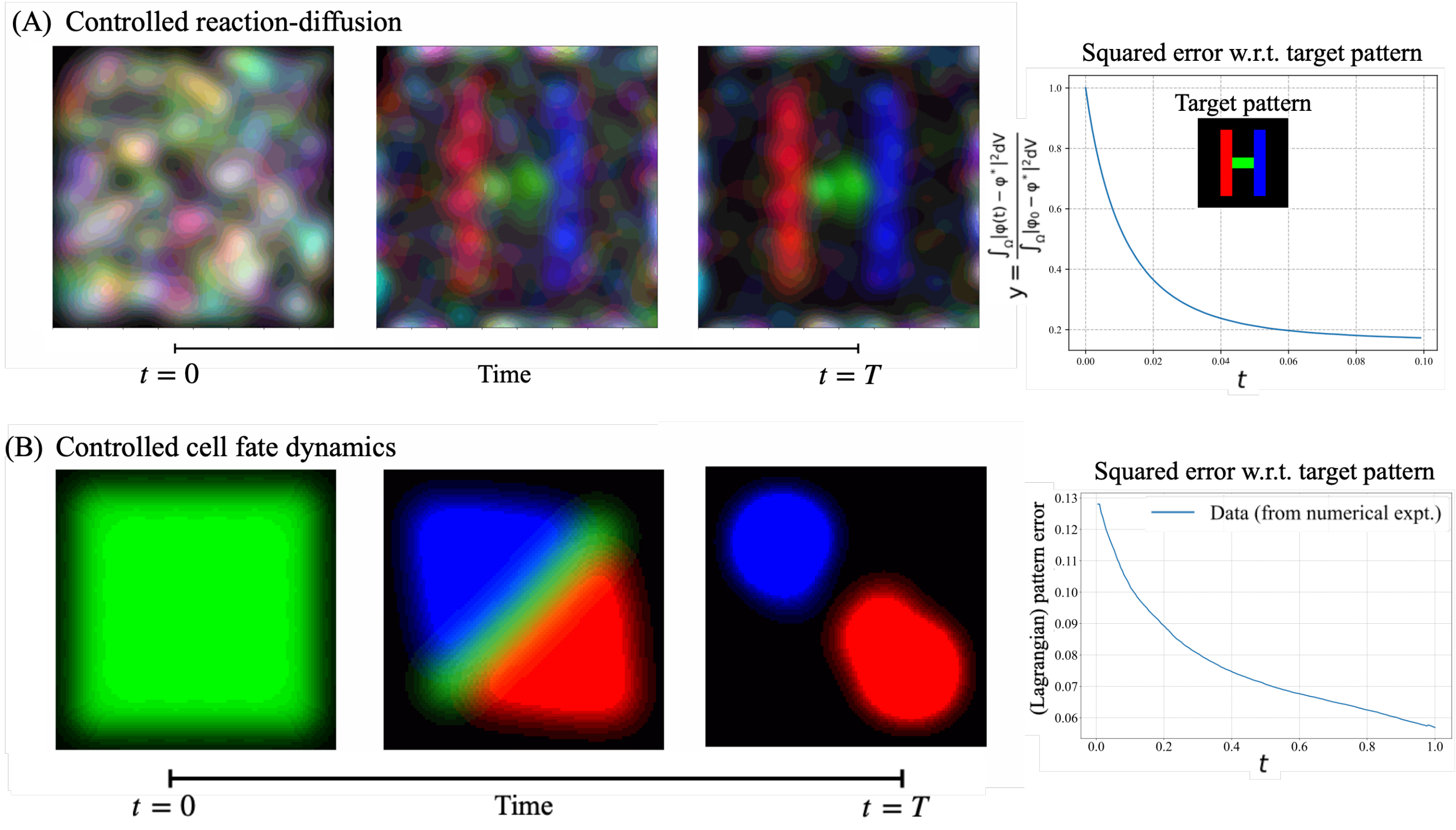}
    \caption{\textbf{Targeted morphogenesis.} {(A) Control of a reaction-diffusion system.} Controlling a reaction-diffusion system comprising three species, from a randomly generated reaction matrix and initial conditions (see SI section S7 for details). The uncontrolled reaction-diffusion system is given by (see (11) and surrounding text) $\frac{\partial \phi_i}{\partial t} = D \Delta \phi_i + f_i(\phi)$, where $\phi(t, {\bf z}) = \left(\phi_1(t, {\bf z}), \phi_2(t, {\bf z}), \phi_3(t, {\bf z}) \right)$ is the vector of concentration of species~$i=1,2,3$ at time~$t$ and~${\bf z} \in \Omega$, $f_i$ is the reaction field for species~$i$ and $D$ is the diffusion constant. The rightmost panel shows that the evolution of the Lagrangian squared error (normalized w.r.t. initial error) relative to target pattern decays appproximately  exponentially (inset shows the $H$-shaped target pattern with the vertical and horizontal bars to be formed by the three individual species, respectively). {(B) Control of cell fate dynamics.} The three panels are the representative intermediate patterns corresponding to the controlled cell fate decision making model (see (12) and surrounding text). The forward uncontrolled equations are given by $\frac{\partial \rho}{\partial t} = \nabla \cdot \left[ \rho \left(\nabla \left( \frac{\delta E}{\delta \rho} \right)\right) \right]$ and $\frac{\partial \phi}{\partial t} - \left[\nabla \left( \frac{\delta E}{\delta \rho} \right) \right] \cdot \nabla \phi =  - \frac{\delta E}{\delta \varphi} + R$ , where $E(\rho, \varphi) = \int_{\Omega} \left( f(\rho) + g(\rho) \left\| \nabla \varphi \right\|^2 \right) d^2 \mathbf{z}$. The figure on the left is the initial pluripotent state, the figure in the middle is the snapshot of the dynamical system at time $T/2$, and the rightmost figure is the snapshot of the dynamical system at time $T$. The rightmost panel shows that the evolution of the Lagrangian pattern error (normalized w.r.t. initial error), relative to target pattern decays appproximately  exponentially.} 
    \label{fig:S-T}
\end{figure}

\noindent \textbf{(A) Control of reaction-diffusion systems.}
Consider a pattern composed of three morphogen fields represented as~$\phi(t,{\bf z}) = (\phi_1(t, {\bf z}),\phi_2(t, {\bf z}), \phi_3(t, {\bf z}))$, 
where $\phi_i(t,\mathbf{z})$ is the concentration of the $i$-th morphogen at a location $\zvec \in \Omega \subset \real^2$ at time $t\in [0, T]$.  We first define an energy functional with a cubic coupling so that the kinetic equations have quadratic nonlinearities, given by
\begin{align*}
    E(\varphi)=\int_{\Omega} [\frac{1}{2}D_i ||\nabla \varphi_i||^2 - M_{ijk}\varphi_i \varphi_j\varphi_k]d^2{\bf z},
\label{var_rd}
\end{align*}
where $i \in (1, 2, 3)$, $D_i$ is the diffusion constant corresponding to the $i$-th morphogen (using the Einstein summation convention in~\eqref{var_rd}), and the term $M_{ijk} \phi_i \phi_j\phi_k$ is a cubic polynomial with $M_{ijk}$ a third order tensor with no assumed symmetries. The passive forward reaction-diffusion dynamics for the $i$-th morphogen is given by $\frac{\partial \phi_i}{\partial t} = - \left(\frac{\delta E}{\delta \varphi} \big|_{\phi}\right)_i =  D_i \Delta \phi_i + M_{ijk} \phi_j \phi_k$. The optimal control problem for the reaction-diffusion system then reads as
\begin{align}
    \min_{{\bf u}, S } 
    ~ &\frac{\gamma_u}{2}  \int_{0}^{T} \int_{\Omega} \left(\sum_{i=1}^3 \phi_i(t,\mathbf{z}) \right) ||{\bf u}(t,\mathbf{z})||^2 d^2\mathbf{z} dt 
    + \frac{\gamma_v}{2} \sum_{i=1}^3 \int_{0}^{T} \int_{\Omega} {S_i^2(t,\mathbf{z})} d^2\mathbf{z} dt + \Psi \left(\phi^{(T)} \right)
    \nonumber \\ & \text{s.t.}~ \frac{\partial \phi_{i}}{\partial t} + \nabla\cdot (\phi_i {\bf u}) = D_i \Delta \phi_i + M_{ijk} \phi_j \phi_k + S_i, ~i = 1,2,3
\end{align}
We use the smoothed particle hydrodynamics-based approach to convert the problem to a Lagrangian framework (see SI section S4 for details), and then deploy our adjoint path-integral  approach~\cite{SS-VK-LM:23} to solve the resulting large system. In our numerical experiments (see SI section S6 for details on discretization), for simplicity, we assume $D_1=D_2=D_3$ and further that $M_{ijk}$ is a third order tensor whose elements are sampled from a normal distribution with mean zero and unit variance. Figure \ref{fig:S-T}A shows the results of an experiment where we steer a initially random pattern of $\phi_i$'s encoded via color e.g. R, G, B to the pattern `H' where the left (right) vertical line is red (blue) and the center horizontal line is green (see Movie4).  \\

\noindent \textbf{(B) Control of cell fate dynamics.}
%
We finally turn to a question inspired by biological morphogenesis wherein the fates and positions of active cells in a multi-cellular tissue are organized spatiotemporally in a reproducible way~\cite{GS-DD-DC:21,  XQ-others:22}. Since cells fates are intimately related to gene-expression patterns within a cell, which are themselves partly steered by their extra-cellular environment, a natural framework requires us to couple the external dynamics of motile cells and their internal states. Indeed, this framework might be relevant to an even broader class of active systems consisting of particles that can have multiple internal states while also capable of moving autonomously, with the states and movements being coupled to each other, e.g. in active matter systems~\cite{MCM-JFJ-SR-TBL-JP-MR-RAS:13}, or traveling waves of differentiation and movement in embryo development \cite{SDT-MV:22}. In a minimal setting, we model a two-dimensional epithelial layer undergoing the process of differentiation associated with the control of cell fate and position starting from a pluripotent state. Consider cells with spatial density function $\rho(t, \mathbf{z})$ and a spatial gene expression profile $\phi(t, \mathbf{z})$ at time~$t$ and $\mathbf{z} \in \Omega$. For simplicity we assume that a particular gene in each cell can take on three values $\lbrace -1, 0, 1 \rbrace$, where the value $0$ corresponds to the undifferentiated pluripotent state and the values ~$-1, 1$ correspond to two distinct differentiated states.  Spatial coarse-graining towards a long wavelength continuum theory allows us to relax this discrete set to a continuum so that the gene expression state  $-1\leq \phi(t, \mathbf{z})\leq 1$. The optimal control problem of creating a non-equilibrium patterned tissue with a specific spatially differentiated phenotype can then be cast as 
\begin{align}
    \min_{{\bf u}, S} 
    ~ &\frac{\gamma_u}{2} \int_{0}^{T} \int_{\Omega} \rho(t, \mathbf{z}) ||{\bf u}(t, \mathbf{z})||^2 d^2 \mathbf{z} dt 
    + \frac{\gamma_v}{2} \int_{0}^{T} \int_{\Omega} \rho(t, \mathbf{z}) S^2(t, \mathbf{z}) d^2 \mathbf{z} dt
    + \Psi(\rho^{(T)}, \phi^{(T)})
    \nonumber \\ & \text{s.t.}~ 
    \begin{cases}
        &\frac{\partial \rho}{\partial t} + \nabla \cdot \left[ \rho \left(- \nabla \left( \frac{\delta E}{\delta \varrho} \right) + \mathbf{u} \right) \right] = 0 \\
        &\frac{\partial \phi}{\partial t} + \left[-\nabla \left( \frac{\delta E}{\delta \varrho} \right) + \mathbf{u} \right] \cdot \nabla \phi =  - \frac{\delta E}{\delta \varphi} + S 
    \end{cases}
    \label{eq:cell_diff}
\end{align}
where the energy functional~$E$ is given by
\begin{align*}
    E(\varrho, \varphi) &= \int_{\Omega} \left( f(\varrho) + g(\varrho) \left\| \nabla \varphi \right\|^2 \right) d^2 \mathbf{z}.
\end{align*}
We note that the intercellular interaction consists of a long-range attraction ($f(\rho)$) and short-range repulsion ($g(\rho)$) between different cell types (See SI section S7 for both the discretized version of the problem and the exact form of $E(\varrho, \varphi)$). The first equation in \eqref{eq:cell_diff} describes the dynamics of the controlled cell population as it progresses due to a combination of internal dynamics and external control (here we assume that total cell number is conserved, even though cell type is not), while the second equation in (14) describes the dynamics of cell fate which is advected at a velocity determined by active cell movement, while simultaneously being changed via reactive terms associated with the effect of the local cell population, as well as external control.  The terminal penalty is given by
\begin{align*}
    \Psi(\varrho, \varphi) &= \frac{\delta_{c}}{4} \int_{\Omega} \prod_{k=1}^2 \left( \varphi - C_k \right)^2 \varrho d^2 \mathbf{z}  + \frac{\delta_\mu}{2} \int_{\Omega} \sum_{k=1}^2  e^{-\frac{(\varphi - C_k)^2}{2\sigma^2}} \left\| \mathbf{z} - \mu^*_k \right\|^2 \varrho d^2 \mathbf{z}  
\end{align*}
where $C_k = \lbrace -1, 1 \rbrace$ characterizes the final differentiated states at spatial location $\mu^*_k$ (where $k=1,2$ corresponding to the differentiated states -1 and 1) and $\sigma$ sets the spatial range of specificity of the morphogen to the gene expression state. The first term in the terminal cost corresponds to differentiation and penalizes undifferentiated (internal) terminal states, while the second term corresponds to segregation and penalizes deviation from (external) target locations for the population. 

Using our SPH approach, we convert the above Eulerian description to a Lagrangian one and then deploy our algorithm for stochastic optimal control (see SI Section S4 for the SPH approach and SI section S7 for the discrete problem formulation). Figure \ref{fig:S-T}B shows the sequential time snapshots of the controlled dynamical system where the cells undergo differentiation from a homogeneous pluripotent state (green) to a differentiated spatially segregated state (see Movie5) as the area of the  differentiated states (i.e red and blue) initially increases and then reaches a steady state in time as cells simultaneously differentiate and migrate to specific target sites (see SI Figure S4).  

\section*{Optimal control as a generative modeling framework}
So far,  we have only considered passive physics given by gradient flows which might seem very restrictive at first. We now show that gradient flows emerge naturally from the framework of optimal control and hence form a necessary and sufficient class for very large class of pattern-forming PDEs. 
%
In the absence of any intrinsic pattern forming capacity, corresponding to (1) with $\mathbf{v} = 0$ and $R = 0$, we first ask how to steer a pattern forming system using a combination of a conserved flux-based steering control  vector field~$\mathbf{u}$ and a non-conserved reaction field~$S$. 
%
%
In terms of the value function given by
\begin{align} \label{eq:pattern_formation_value_func}
\begin{aligned}
    \mathcal{F}(t, \varphi)=  \min_{\mathbf{u}} ~&\frac{\gamma_{\ctrl}}{2} \int_t^T \int_\Omega m(\phi(\tau, {\bf z})) \left\| \mathbf{u}(\tau, {\bf z}) \right\|^2 d^2 \mathbf{z} d\tau + \frac{\gamma_{\vctrl}}{2} \int_t^T \int_\Omega S^2(\tau, {\bf z}) ~d^2 \mathbf{z} d\tau + \mathcal{J}\left(\phi^{(T)} \right), \\ &\text{s.t.} \begin{cases} \frac{\partial \phi}{\partial \tau}(\tau, {\bf z}) + \nabla \cdot \left( m(\phi(\tau, {\bf z})) \mathbf{u}(\tau, {\bf z}) \right) = S(\tau, {\bf z}) \\
 \phi(t, \mathbf{z}) = \varphi(\mathbf{z}) \end{cases}
\end{aligned}
\end{align}
we ask how to achieve a prescribed target pattern, but now using a relaxation of the terminal constraint to a smooth version given by~\eqref{eq:opt_ctrl_pattern_formation_terminal_constraints}, replacing it with a patterning objective functional $\mathcal{J}(\varphi) = \int_\Omega [\Psi(\varphi(\mathbf{z})) + \frac{\epsilon}{2} \left\| 
 \nabla \varphi(\mathbf{z}) \right\|^2]d^2 \mathbf{z}$. We note that here $\mathcal{F}(t, \varphi)$ is the value function corresponding to the problem of optimally steering the system over the infinite-dimensional pattern space from a pattern~$\varphi$ at time~$t$ to minimize the patterning objective~$\mathcal{J}$. 
Using the calculus of variations~\cite{RB:57, FT:10}, it can be shown (see Section S8 for details) that the value function~$\mathcal{F}(t, \varphi)$ satisfies a nonlinear PDE
\begin{align} \label{eq:HJB}
\begin{aligned}
    \frac{\partial \mathcal{F}}{\partial t} + \frac{1}{2\gamma_{\ctrl}} \int_\Omega  m(\varphi) \left\| \nabla \left( \frac{\delta \mathcal{F}}{\delta \varphi} \right) \right\|^2 d^2\mathbf{z} + \frac{1}{2 \gamma_{\vctrl}} \int_\Omega \left( \frac{\delta \mathcal{F}}{\delta \varphi} \right)^2 d^2\mathbf{z} = 0& \quad \text{holds~for~any~} t, \varphi \\ 
    \mathcal{F}(T,\varphi) = \int_{\Omega} \left[\Psi(\varphi({\bf z})) + \frac{\epsilon}{2} \left\| \nabla \varphi(\mathbf{z}) \right\|^2 \right] d^2 \mathbf{z}.&
\end{aligned}
\end{align}
which can be viewed as a generalization of the Hamilton-Jacobi-Bellman (HJB) equation to infinite-dimensional pattern spaces \cite{FT:10}. Furthermore, the optimal control vector field $\mathbf{u} = - \frac{1}{\gamma_{\ctrl}} \nabla \left( \frac{\delta \mathcal{F}}{\delta \varphi} \right)$ and the optimal control reaction field $S = - \frac{1}{\gamma_{\vctrl}} \frac{\delta \mathcal{F}}{\delta \varphi}$ (see Section S8 for details).  

How can we transition from the above {\it tabula-rasa} model that has no intrinsic pattern forming capacity, but is fully controlled and gradually add intrinsic pattern formation capacity, and potentially reduce the amount of control? A natural way is to generate complex pattern forming functionality in a sequence of incremental steps associated with a sequence of simpler value functions, e.g. $\mathcal{F}_i,~ i=0,1,2,...M $  representing a sequence of $M$ accessible way points encoded in a sequence of patterning costs $\left \lbrace \Psi_k(\varphi) \right \rbrace_{k=1}^M$. Then, starting without intrinsic potential, i.e, ${\bf v}=0$, $R=0$, we can introduce a rudimentary pattern forming functionality via gradient fields $\mathbf{v} = - \frac{1}{\gamma_{\ctrl}} \nabla \left( \frac{\delta \mathcal{F}_0}{\delta \varphi} \right)$, $R = - \frac{1}{\gamma_{\vctrl}} \frac{\delta \mathcal{F}_0}{\delta \varphi}$. Bootstrapping from this simple pattern forming system with intrinsic pattern forming capacity given by $\mathbf{v}, R$, the pattern forming functionality can be expanded by a gradual addition of pattern forming capacity into the control objective.
This provides a generative sequence that can result in patterns with increasing complexity, defined in terms of a sequence of ``terminal" costs associated with way points, providing among other things, an interpretation of the evolution of complexity over very long time scales in biological pattern forming systems, learning of pattern generation functionality from a demonstration by a controller which provides a sequence of simple intermediate goals, etc.


%

\section*{Discussion}

Our study provides a framework to formulate and solve inverse design problems associated with pattern control in non-equilibrium physical, chemical and biological systems.  We show how minimizing control costs and terminal target-driven goals converts predictive theories for long wavelength pattern formation-  themselves grounded in conservation laws and symmetry/broken symmetry-   into problems in stochastic optimal control theory for spatio-temporal fields.   We note that the ``Hamiltonian bridge'' formalism based on optimal control naturally includes the notion of optimal interpolation building on generative models, such as those based on
normalizing flows~\cite{EGT-EVE:10,EGT-CVT:13,DR-SM:15} and diffusion models~\cite{JSD-EW-NM-SG:15,YS-JSD-DPK-AK-SE-BP:20}. By accounting for physical priors embodied in terms of pattern formation dynamics, it also generalizes the classical optimal transport perspective~\cite{JDB-YB:00} which do not account for these dynamical constraints imposed by physics. 

An Eulerian-Lagrangian mapping allows us to convert the field theory into a scale-dependent approximate many-body control problem that can be solved numerically using continuous time back-propagation methods originally devised for finite-dimensional systems \cite{RTQC-YR-JB-DKD:18,XL-TKL-RTQC-DD:20,SS-VK-LM:23}. Our numerical experiments demonstrate the effectiveness of our physics-driven generative framework for a range of pattern forming systems that include phase separation with and without a conserved order parameter, and morphogenetic processes involving reaction-diffusion and activity. Our framework thus suggests a mechanism for how we can drive pattern-forming capacity in a system that starts out having none, using optimal control to create a series of way points (rather than a single terminal cost) in a closed-loop feedback system to drive forward physics  step by step.
 

Finally, we close with a view of what might be next. Traditional theory-rich approaches to pattern formation have relied on physical intuition and domain expertise to create coarse-grained phenomenological models ~\cite{GTR-CBM-TS-SYS:06}, combined with analytical and numerical techniques. Recent data-rich advances in generative models have opened up new data-driven avenues for studying and manipulating patterns~\cite{HY:22}. Our approach naturally suggests an inference engine for generating and controlling patterns using physics-based priors (when available) along with experiments (to guide parameter choices) using optimal control and forward-backward algorithms that can be efficiently deployed across scales and~systems.


\bibliographystyle{unsrt}
\bibliography{references}

\renewcommand{\thesection}{S\arabic{section}}
\renewcommand{\theequation}{S.\arabic{equation}}
\renewcommand{\thefigure}{S\arabic{figure}}

\titleformat{\section}
  {\huge\bfseries\filcenter}{\thesection}{1em}{}

\section*{\underline{Supplementary information}}

\titleformat{\section}
  {\LARGE}{\thesection}{1em}{}

\vspace{0.2in}

\section*{S0~~~Details of numerical experiments}
\begin{itemize}

     \item \textbf{Controlling Allen-Cahn (Movie1):} Controlling a square lattice chequered board pattern of blue and white colors governed by the Allen-Cahn Equation. The uncontrolled Allen-Cahn equation is given by $\frac{\partial \phi}{\partial t} =R$, where $R$ is the reaction field given by $R=- \frac{\delta E }{\delta \varphi}$, where $E(\varphi)=\int_{\Omega} \left[ U(\varphi(\mathbf{x})) + \frac{\epsilon}{2} \left\| \nabla \varphi(\mathbf{x}) \right\|^2 \right] d^2 \mathbf{z}$ and  $U(\varphi)=(1-\varphi^2)^2$, $\epsilon=10^{-3}$. The target pattern is `H' in white colors in the center surrounded by blue. The simulation details are as follows: $m(\phi)=0, {\bf v}=0, \gamma_{u}=0, R=-\frac{\partial E}{\partial \varphi}$ where $E(\varphi)=\int_{\Omega} \left[ (1-\varphi^2)^2 + \frac{\epsilon}{2} \left\| \nabla \varphi(\mathbf{x}) \right\|^2 \right] d^2 \mathbf{z}, \epsilon=0.001, \gamma_v=1, T=0.01, dt=0.001$. The initial pattern was a random chequered pattern where 25 static lagrangian particles where initialized on a $5\times 5$ lattice in a box size of approximately 3 units. The target pattern was `H'. 
     
    \item \textbf{Controlling Cahn-Hilliard (Movie2):} Controlling a random distribution of blue and white colors governed by the Cahn-Hilliard equation.  The uncontrolled Cahn-Hilliard equation is given by $\frac{\partial \phi}{\partial t} + \nabla \cdot \left( m(\phi)  \mathbf{v} \right)=0$, where $\phi$ is the time-dependent phase field, $m(\varphi)=m=1$ is the mobility, and $\mathbf{v}$ is the known vector field given by $\mathbf{v}=-\nabla \frac{\delta E }{\delta \varphi}$, where $E(\varphi)=\int_{\Omega} \left[ U(\varphi(\mathbf{x})) + \frac{\epsilon}{2} \left\| \nabla \varphi(\mathbf{x}) \right\|^2 \right] d^2 \mathbf{z}$ and  $U(\varphi)=(1-\varphi^2)^2$, $\epsilon=10^{-3}$. The target pattern was `H'. The simulation details are as follows: $m(\phi)=1, {\bf v}=-\frac{\partial E}{\partial \varphi}$ where $E(\varphi)=\int_{\Omega} \left[ (1-\varphi^2)^2 + \frac{\epsilon}{2} \left\| \nabla \varphi(\mathbf{x}) \right\|^2 \right] d^2 \mathbf{z}, \epsilon=0.001, \gamma_{u}=1, R=0, \gamma_v=0, T=0.01, dt=0.001$. The initial condition by sampling 25 movable lagrangian particles using a uniform distribution in a square box of size approximately 3 units. The target eulerian pattern was `H'.

    \item \textbf{Controlling Droplets (Movie3):} The uncontrolled thin film equation is given by $\frac{\partial \phi}{\partial t} + \nabla \cdot \left( m(\phi)  \mathbf{v} \right)=0$, where $\phi$ is the time-dependent height function of the droplet, $m(\varphi)=\varphi^3$ is the mobility, and $\mathbf{v}$ is the known vector field given by $\mathbf{v}=-\nabla \frac{\delta E }{\delta \varphi}$, where $E(\varphi)=\int_{\Omega} \left[ U(\varphi(\mathbf{x})) + \frac{\epsilon}{2} \left\| \nabla \varphi(\mathbf{x}) \right\|^2 \right] d^2 \mathbf{z}$ and  $\frac{dU}{d\varphi}=\varphi e^{-\varphi^2}(\varphi^2-\eta_1)$, where $\epsilon=10^{-3}$ and $\eta=10$. The simulation details are as follows: $m(\phi)=\phi^3, {\bf v}=-\frac{\partial E}{\partial \varphi}$ where $E(\varphi)=\int_{\Omega} \left[ U(\varphi(\mathbf{x})) + \frac{\epsilon}{2} \left\| \nabla \varphi(\mathbf{x}) \right\|^2 \right] d^2 \mathbf{z}$ and  $\frac{dU}{d\varphi}=\varphi e^{-\varphi^2}(\varphi^2-\eta_1)$, where $\epsilon=10^{-3}$ and $\eta=10$, $\gamma_{u}=1, R=0, \gamma_v=0, T=0.01, dt=0.001$. The initial condition by sampling 25 movable lagrangian particles using a uniform distribution in a square box of size approximately 3 units. The target eulerian pattern was a $5\times5$ square lattice.

    \item \textbf{Controlling Reaction-Diffusion system (Movie4):} The uncontrolled reaction-diffusion system is given by $\frac{\partial \phi_i}{\partial t} = D \Delta \phi_i + f_i(\phi)$, where $\phi(t, {\bf z}) = \left(\phi_1(t, {\bf z}), \phi_2(t, {\bf z}), \phi_3(t, {\bf z}) \right)$ is the vector of concentration of species~$i=1,2,3$ at time~$t$ and~${\bf z} \in \Omega$, $f_i$ is the reaction field for species~$i$ and $D$ is the diffusion constant. See section S6 for the discrete formulation of the problem. The simulation details are as follows: $D=1$, $M_{ijk}$ are three tensor elements of which are sampled form mean zero and unit variance, $\gamma_u=\gamma_v=1, T=0.01, dt=0.001$. The initial distribution is random pattern of the three fields which were initialized on 500 lagrangian particles sampled uniformly in a square box of approximately 3 units and the target is pattern `H' where the two vertical bars are blue and red, and the horizontal bar is green. 
    
    \item \textbf{Controlling Cell-fate (Movie5):} The forward uncontrolled equations are given by $\frac{\partial \rho}{\partial t} = \nabla \cdot \left[ \rho \left(\nabla \left( \frac{\delta E}{\delta \varrho} \right)\right) \right]$ and $\frac{\partial \phi}{\partial t} - \left[\nabla \left( \frac{\delta E}{\delta \varrho} \right) \right] \cdot \nabla \phi =  - \frac{\delta F}{\delta \varphi} + R$ , where $E(\varrho, \varphi) = \int_{\Omega} \left( f(\varrho) + g(\varrho) \left\| \nabla \varphi \right\|^2 \right) d^2 \mathbf{z}$. The simulation details are as follows: $\gamma_u=\gamma_v=1,\delta_c=1,~\delta_{\mu}=1,~C_k=[-1,1],~\sigma=1, ~\mu_k=[[0,5],[5,0]], T=1,~dt =0.01$. The initial condition was prepared by initializing 25 lagrangian particles on a $5\times 5$ latticle all with the same $C=0$ (i.e they have the same state). The target was to minimize the terminal state defined on the lagrangian configuration given by $\Psi(\varrho, \varphi) = \frac{\delta_{c}}{4} \int_{\Omega} \prod_{k=1}^2 \left( \varphi - C_k \right)^2 \varrho d^2 \mathbf{z}  + \frac{\delta_\mu}{2} \int_{\Omega} \sum_{k=1}^2  e^{-\frac{(\varphi - C_k)^2}{2\sigma^2}} \left\| \mathbf{z} - \mu^*_k \right\|^2 \varrho d^2 \mathbf{z}$ (see section S7 for the discrete version).
\end{itemize}

\section{Mathematical notation}
\begin{itemize}
    \item $\Omega \in \real^2$ is an unbounded planar Euclidean domain.
    \item $\mathbf{z}, d^2\mathbf{z}:$ Eulerian co-ordinate and the area element on $\Omega$.
    
    \item ${\phi:}$ Time-varying pattern field, $\phi \in \real_{+} \times \mathcal{G}$, where $\mathcal{G} = \lbrace \varphi:  \Omega \rightarrow \real^{Q} \rbrace \subseteq L^2(\Omega)$ is the space of patterns. In most of the examples, except the cell fate control and  reaction-diffusion (in the SI section S7 and S8), we consider a scalar field (i.e $Q=1$).

    \item $\varphi$: a generic pattern, point in pattern space~$\mathcal{G}$.

    \item $m({\varphi}):$ Mobility field ($m : \mathcal{G} \rightarrow \real$).

    \item $\mathbf{v}:$ Known passive vector field ($\mathbf{v} : \real_{+} \times \Omega \rightarrow \real^2$ is such that $\mathbf{v} \in \real_{+} \times L^2(\Omega; \real^2)$). 

    \item $\mathbf{u}:$ Unknown control vector field ($\mathbf{u}:\real_{+} \times \Omega \rightarrow \real^2$ is such that $\mathbf{u} \in \real_{+} \times L^2(\Omega; \real^2)$).

    \item $R:$ Known reaction field ($R:\real_{+} \times \Omega \rightarrow \real^{Q}$ is such that $R \in \real_{+} \times L^2(\Omega; \real^Q)$).

    \item $S:$ Unknown control reaction field ($S:\real_{+} \times \Omega \rightarrow \real^Q$ is such that $S \in \real_{+} \times L^2(\Omega; \real^Q)$)

    \item $C_{\gamma_{\ctrl}, \gamma_{\vctrl}}(\varphi, \mathbf{u}, S):$ Instantaneous steering cost, where $C_{\gamma_{\ctrl}, \gamma_{\vctrl}}: \mathcal{G} \times L^2(\Omega; \real^Q) \times L^2(\Omega; \real^Q)$ and $\gamma_{\ctrl}$ and $\gamma_{\vctrl}$ are the real-valued parameters weighting the cost controlling vector and reaction fields respectively.


    \item $\phi_0,\phi^*$: Initial and target pattern field.

    \item $\Lambda \in \real_{+} \times \bar{\mathcal{G}}$ is the co-state corresponding to state~$\phi$, where $\bar{\mathcal{G}}$ is the dual space of $\mathcal{G}$.

    \item $\lambda$: any point in the dual space~$\bar{\mathcal{G}}$.

    \item $H(\varphi, \lambda):$ Hamiltonian density defined over fields $\varphi$ and $\lambda$.

    \item $H_0 \left(\varphi, \lambda \right):$ Nominal Hamiltonian.

     \item $\mathcal{F}(t, \varphi):$ The optimal control value function satisfying the Hamilton-Jacobi-Bellman equation (Eqn. (13-14) in the main text).

      \item $E_1(\varphi), E_2(\varphi):$ Energy functionals.

     \item $U(\varphi):$ Potential energy density.

    \item $\xvec(t):$ The positional configuration of the $N$ particle system at time $t$ with $\xvec(t) \in \real^{Nd_x}$. Here, $d_x$ is the ambient dimensionality which in the present study is two. The velocity $\frac{d\xvec(t)}{dt}=\dot{\xvec}(t)$.

    \item $\phivec(t):$ The state configuration of the $N$ particle system at time $t$ with $\phivec(t) \in \real^{Nd_\phi}$.

    \item $\muvec(t, \xvec, \phivec), \nuvec(t, \xvec, \phivec):$ Known drift vector fields corresponding to $\xvec(t)$ and $\phivec(t)$ respectively.

    \item $\sqrt{2D_\xi} \xivec(t),\sqrt{2D_\eta} \etavec(t):$  Diffusion terms corresponding to $\xvec(t)$ and $\phivec(t)$ respectively where $D_\xi$ and $D_\eta$ dictate the strength of fluctuations and $ \xivec(t)$ and $\etavec(t)$ correspond to fluctuations with gaussian-white noise statistics.

    \item $\ctrlvec (t), \vvec(t):$ Controls which enable the steering of the configuration-state pair~$(\xvec, \phivec)$.

    \item $p(t,\xvec,\phivec):$ Probability density for the state pair to be at $(\xvec, \phivec)$ at time $t$. The probability density satisfies the property $\int_{\real^{N d_x}\times \real^{N d_{\phi}}} p(t,\xvec,\phivec) d\xvec d\phivec =1$, $\forall t\in [0,T]$.

    \item  $\varphi(t,s,\zvec):$ Coarse-grained Eulerian description the state~$(\xvec, \phivec)$ via a kernel $K$ as follows $\varphi(t,s,\zvec)=\sum_{i=1}^N K(s, \zvec - \xvec_i(t)) \phi_i(t)$ where $s$ is the scale parameter and $z \in \real^{d_x}$ is the Eulerian coordinate.

    \item $D(p_T, p^*):$ Distance between the terminal probability density $p(T,\xvec,\phivec)$ and the desired probability density $p^*$.

\end{itemize}

\section{Derivation of the Hamiltonian bridge} \label{sec:si_hamiltonian_bridge}
The active pattern control problem of steering a pattern~$\phi_0$ at time $t=0$ to a target pattern~$\varphi^*$ at time~$t=T$ is given by
\begin{align} \label{eq:opt_ctrl_pattern_formation_terminal_constraints}
    \min_{\mathbf{u}, S} ~  \int_0^T C_{\gamma_{\ctrl}, \gamma_{\vctrl}}(t, \phi, \mathbf{u}, S) dt \quad \text{s.t.} \quad \begin{cases} \frac{\partial \phi}{\partial t} + \nabla \cdot \left( m(\phi) ( \mathbf{v} + \mathbf{u} ) \right) = R + S \\
    \phi(0, {\bf z}) = \phi_0({\bf z}), ~\phi(T,{\bf z}) = \phi^*({\bf z})
    \end{cases}
\end{align}
Constructing an action functional for the above problem, given by
\begin{align} \label{eq:optimal_ctrl_action}
    \begin{aligned}
    \mathcal{A} &= \int_0^T C_{\gamma_{\ctrl}, \gamma_{\vctrl}}(t, \phi, \mathbf{u}, S) dt
                    - \int_0^T \int_\Omega \Lambda \left( \frac{\partial \phi}{\partial t} + \nabla \cdot \left( m(\phi) ( \mathbf{v} + \mathbf{u} ) \right) - R - S \right) d^2\mathbf{z} dt \\
                &= \int_\Omega \Lambda_0 \phi_0 d^2 \mathbf{z} - \int_\Omega \Lambda_T \phi_T d^2 \mathbf{z} + \int_0^T C_{\gamma_{\ctrl}, \gamma_{\vctrl}}(t, \phi, \mathbf{u}, S) dt + \int_0^T \int_\Omega \phi \frac{\partial \Lambda}{\partial t} d^2 \mathbf{z} dt \\ 
                &\qquad + \int_0^T \int_\Omega m(\phi) \nabla \Lambda \cdot (\mathbf{v}+\mathbf{u}) ~d^2 \mathbf{z} dt
                 +  \int_0^T \int_\Omega \Lambda (R + S) ~d^2 \mathbf{z} dt
    \end{aligned}
\end{align}
Note that the solution to the optimal control problem~\eqref{eq:opt_ctrl_pattern_formation_terminal_constraints} is obtained as the solution to the following $\min-\max$ problem involving the action functional~$\mathcal{A}$
\begin{align*}
    \min_{\phi, \mathbf{u}, S} \max_{\Lambda} ~\mathcal{A}(\phi, \mathbf{u}, S)
\end{align*}
For the instantaneous steering cost given by
\begin{align*}
C_{\gamma_{\ctrl}, \gamma_{\vctrl}}(t, \phi, \mathbf{u}, S) = \frac{\gamma_{\ctrl}}{2} \int_\Omega m(\phi(t, \mathbf{z})) \left\| \mathbf{u}(t, {\bf z}) \right\|^2 d^2 \mathbf{z} + \frac{\gamma_{\vctrl}}{2} \int_\Omega  \left( S(t, {\bf z}) \right)^2  d^2 \mathbf{z}    
\end{align*}
we note that the optimal control vector field~$\mathbf{u}^*$ and reaction field~$S^*$, obtained from minimizing the action~$\mathcal{A}$ above first with respect to~$\mathbf{u}$, $S$ (note that the order of the minimization and maximization can be reversed, since the action functional is convex in $\phi, \mathbf{u}$ and linear, hence concave in $\Lambda$) are given by
\begin{align*}
    \mathbf{u}^*(t, \mathbf{z}) = - \frac{1}{\gamma_{\ctrl}} \nabla \Lambda(t, \mathbf{z}), \qquad  S^*(t, \mathbf{z}) = - \frac{1}{\gamma_{\vctrl}} \Lambda(t, \mathbf{z})
\end{align*}
Now, constructing a \emph{control Hamiltonian} functional~$H$ as follows
\begin{align} \label{eq:hamiltonian}
    H \left(\varphi, \lambda \right) = \int_\Omega \left[ m(\varphi) \nabla \lambda \cdot \mathbf{v} + \lambda R - \frac{m(\varphi)}{2\gamma_{\ctrl}} \left\| \nabla \lambda \right\|^2 - \frac{\lambda^2}{2\gamma_{\vctrl}} \right] d^2 \mathbf{z}
\end{align}
where $(\varphi, \lambda)$ is any pattern--co-state pair, we note that the optimal pattern evolution (i.e., the solution to Problem~\eqref{eq:opt_ctrl_pattern_formation_terminal_constraints} obtained as the stationary trajectory of the action functional~$\mathcal{A}$ in \eqref{eq:optimal_ctrl_action}) can be expressed in the form of a Hamiltonian system 
\begin{align} \label{eq:hamilton_bridge}
    \frac{\partial \phi}{\partial t} = \left \lbrace \phi, H \right \rbrace, \qquad
    \frac{\partial \Lambda}{\partial t} = \left \lbrace \Lambda, H \right \rbrace
\end{align}
where $\Lambda$ is the co-state corresponding to~$\phi$, $\lbrace \cdot , \cdot \rbrace$ above is the Poisson bracket, such that $\left \lbrace \phi, H \right \rbrace = \frac{\delta H}{\delta \lambda}$ and $\left \lbrace \Lambda, H \right \rbrace = -\frac{\delta H}{\delta \varphi}$. Furthermore, for a state--co-state trajectory pair~$(\phi, \Lambda)$ that is a solution to the Hamiltonian system~\eqref{eq:hamilton_bridge}, what remains of the $\min-\max$ problem involving the action functional~$A$ is the maximization with respect to~$\Lambda$. Therefore, the solution to the optimal control problem~\eqref{eq:opt_ctrl_pattern_formation_terminal_constraints} is equivalently obtained as the solution to the following maximization problem
\begin{align*}
    \max_{\Lambda_0} ~\int_\Omega \Lambda_0 \phi_0 ~d^2 \mathbf{z} - \int_\Omega \Lambda_T \phi^* ~d^2 \mathbf{z} ,
    \quad \text{s.t.} \quad 
    \frac{\partial \phi}{\partial t} = \left \lbrace \phi, H \right \rbrace, ~
    \frac{\partial \Lambda}{\partial t} = \left \lbrace \Lambda, H \right \rbrace
\end{align*}
where $\phi(0, {\bf z}) = \phi_0({\bf z})$, $\Lambda(0, {\bf z}) = \Lambda_0({\bf z})$ and $\Lambda(T, {\bf z}) = \Lambda_T({\bf z})$. 

\section{Geometry of interpolation}
\subsection{Model A: Allen-Cahn}
The Hamiltonian bridge formalism for Model~A corresponds to the case of the vector field~$\mathbf{v}=0$, where we set the control vector field~$\mathbf{u}=0$ and optimally steer the passive pattern forming system from an initial~$\phi_0$ at time~$t=0$ to a target $\phi^*$ at time~$t=T$. The optimal control reaction field $S$ is obtained as the solution to the following problem
\begin{align*}
    \min_{S} &\frac{\gamma_{\vctrl}}{2} \int_0^T \int_\Omega S^2(t, \mathbf{z}) ~d^2\mathbf{z} dt \\
    &\text{s.t.}
    \begin{cases}
        \frac{\partial \phi}{\partial t} = R + S \\
        \phi(0, \mathbf{z}) = \phi_0(\mathbf{z}), ~ \phi(T,\mathbf{z}) = \phi^*(\mathbf{z})
    \end{cases}
\end{align*}
Following the steps from the derivation of the Hamiltonian bridge in Section~S2 earlier, the optimal control reaction field is given by $S(t, \mathbf{z}) = - \frac{1}{\gamma_{\vctrl}} \Lambda(t, \mathbf{z})$, where $\frac{\partial \Lambda}{\partial t} = \lbrace \Lambda, H \rbrace = 0$ (note that since $\mathbf{v}, \mathbf{u} = 0$ in the Hamiltonian~\eqref{eq:hamiltonian}, the Poisson bracket $\lbrace \Lambda, H \rbrace = 0$). Therefore, the optimal control reaction field is time-invariant, while the state--co-state system $(\Phi, \Lambda)$ satisfies the terminal conditions $\phi(0, \mathbf{z}) = \phi_0(\mathbf{z}), ~ \phi(T,\mathbf{z}) = \phi^*(\mathbf{z})$. We can therefore express the optimal pattern evolution explicitly in this case, as follows
\begin{align*}
    \phi(t, \mathbf{z}) = \phi_0(\mathbf{z}) + \int_0^t R(\tau, \mathbf{z}) d\tau + t S(\mathbf{z}), \qquad \phi(0, \mathbf{z}) = \phi_0(\mathbf{z}), ~ \phi(T,\mathbf{z}) = \phi^*(\mathbf{z})
\end{align*}
Solving for $S$ using the terminal conditions, we obtain
\begin{align*}
    S(\mathbf{z}) = \frac{1}{T} \left( \phi^*(\mathbf{z}) - \phi_0(\mathbf{z}) - \int_0^T R(\tau, \mathbf{z}) d\tau \right)
\end{align*}
Substituting the above in the expression for~$\phi(t,\mathbf{z})$, we obtain the optimal interpolant~$\phi$, given by
\begin{align} \label{eq:pure_non-conservative_ctrl_exp_solution}
    \phi(t, \mathbf{z}) = \phi_0(\mathbf{z}) + \frac{t}{T} \left( \phi^*(t, \mathbf{z}) - \phi_0(t, \mathbf{z}) \right) + \int_0^t R(\tau,\mathbf{z}) d\tau - \frac{t}{T} \int_0^T R(\tau,\mathbf{z}) d\tau. 
\end{align}
\subsection{Model B: Cahn-Hilliard}
The Hamiltonian bridge formalism in the previous section allows us to define a discrepancy measure $D_{\gamma_{\ctrl}, \gamma_{\vctrl}}$ between patterns, wherein $D_{\gamma_{\ctrl}, \gamma_{\vctrl}}(\phi_0, \phi^*)$ between patterns~$\phi_0$ and~$\phi^*$ (parameterized by $\gamma_{\ctrl}, \gamma_{\vctrl}$) is given by the optimal value in the above problem. Furthermore, when $D_{\gamma_{\ctrl}, \gamma_{\vctrl}}$ satisfies the properties of a distance function, we obtain a geometry in pattern space, wherein the Hamiltonian bridge corresponds to the geodesic between any two patterns $\phi_0$ and $\phi^*$. Following the theoretical formalism of the Hamiltonian bridge, we now seek to better understand the geometry of the optimal pattern evolution. To this end, we consider the case of flux control (i.e., $R=0$ and $\int_\Omega \phi_0 d^2 \mathbf{z} = \int_\Omega \phi^* d^2 \mathbf{z}$) whereby the target pattern is achieved by steering the pattern evolution by a control vector field~$\mathbf{u}$, and the notion of geometry of pattern evolution reduces to looking at the transport paths corresponding to the evolution. We begin with an analysis of the geometry of the optimal pattern evolution under a nominal Hamiltonian (the case $\mathbf{v} = 0$ and penalty $L(\varphi) = 0$) and mobility $m(\varphi) = \varphi$, and later probe the consequences of adding an intrinsic physics to the nominal Hamiltonian via a perturbation analysis, to investigate how adding additional physics changes the geometry of pattern manifold. The nominal Hamiltonian for this case of conserved pattern evolution is given by
\begin{align*}
    H_0 \left(\varphi, \lambda \right) =  - \frac{1}{2\gamma_{\ctrl}} \int_\Omega \varphi \left\| \nabla \lambda \right\|^2 d^2 \mathbf{z} 
\end{align*}
Furthermore, the adjoint equation (i.e., $\partial \Lambda/\partial t = - \delta H_0/\delta \varphi$) is given by
\begin{align*}
    \frac{\partial \Lambda}{\partial t} - \frac{1}{2\gamma_{\ctrl}} \left\| \nabla \Lambda \right\|^2 = 0
\end{align*}
Taking the gradient above, we obtain
\begin{align*}
    \frac{\partial \nabla \Lambda}{\partial t} - \frac{1}{\gamma_{\ctrl}} \nabla^2 \Lambda \nabla \Lambda = 0
\end{align*}
Along the flow $\frac{d}{dt}\mathbf{x}(t) = - \frac{1}{\gamma_{\ctrl}} \nabla \Lambda(t, \mathbf{x}(t))$ with $\mathbf{x}(0) = \mathbf{x}_0$, we obtain
\begin{align*}
    \frac{D}{Dt} \nabla \Lambda(t, \mathbf{x}(t)) = 0
\end{align*}
and it follows that $\nabla \Lambda(t, \mathbf{x}(t)) = \nabla \Lambda(0, \mathbf{x}_0) = \nabla \Lambda_0(\mathbf{x}_0)$. Therefore, we get that the optimal pattern evolution in the nominal case is carried out along straight line paths given by $\mathbf{x}(t) = \mathbf{x}_0 - \frac{t}{\gamma_{\ctrl}} \nabla \Lambda_0(\mathbf{x}_0)$. Furthermore, we can compute $\Lambda_0$ from the terminal constraints~$\varphi_0$ and~$\varphi^*$ on the pattern evolution. We see that the optimal pattern evolution is given by
\begin{align*}
    \frac{\partial \phi}{\partial t} + \nabla \phi \cdot \left( - \frac{1}{ \gamma_{\ctrl}} \nabla \Lambda \right) - \frac{\phi}{\gamma_{\ctrl}} \Delta \Lambda = 0
\end{align*}
so that along the characteristics of the flow $\frac{d}{dt}\mathbf{x}(t) = - \frac{1}{\gamma_{\ctrl}} \nabla \Lambda(t, \mathbf{x}(t)) = - \frac{1}{\gamma_{\ctrl}} \nabla \Lambda_0(\mathbf{x}_0)$, we obtain
\begin{align*}
    \frac{1}{\phi(t, \mathbf{x}(t))} \frac{D}{Dt} \phi(t, \mathbf{x}(t)) 
    = \frac{1}{\gamma_{\ctrl}} \Delta \Lambda(t, \mathbf{x}(t))
\end{align*}
and by integrating the above from $t=0$ to $t=T$, we obtain the governing equation for $\Lambda_0$
\begin{align*}
\log \left( \frac{\phi^* \left(\mathbf{x}_0 - \frac{T}{\gamma_{\ctrl}} \nabla \Lambda_0(\mathbf{x}_0) \right)}{\phi_0(\mathbf{x}_0)} \right) = \frac{T}{\gamma_{\ctrl}} \Delta \Lambda_0(\mathbf{x}_0) 
\end{align*}
In the presence of intrinsic physics, described by a Hamiltonian $H_1\left(\varphi, \lambda \right) = \int_\Omega \left( - \varphi \nabla \lambda \cdot \nabla V \right)d^2 \mathbf{z}$, where $V = V(\mathbf{z})$ is the potential, the flow corresponding to the optimal pattern evolution under the total Hamiltonian~$H = H_0 + H_1$ is given by
\begin{align*}
    \frac{\partial \Lambda}{\partial t} - \frac{1}{2\gamma_{\ctrl}} \left\| \nabla \Lambda \right\|^2 - \nabla \Lambda \cdot \nabla V = 0
\end{align*}
Taking the gradient above, we obtain 
\begin{align*}
    \frac{\partial \nabla \Lambda}{\partial t} + \nabla^2 \Lambda \left( - \frac{1}{\gamma_{\ctrl}} \nabla \Lambda - \nabla V \right) - \nabla^2 V \nabla \Lambda = 0
\end{align*}
Along the flow $\frac{d}{dt}\mathbf{x}(t) = - \frac{1}{\gamma_{\ctrl}} \nabla \Lambda(t, \mathbf{x}(t)) - \nabla V$ with $\mathbf{x}(0) = \mathbf{x}_0$, we obtain
\begin{align*}
    \frac{D}{Dt} \nabla \Lambda(t, \mathbf{x}(t)) - \nabla^2 V(\mathbf{x}(t)) \nabla \Lambda(t, \mathbf{x}(t)) = 0
\end{align*}
Solving the above we get
\begin{align*}
    \nabla \Lambda(t,\mathbf{x}(t)) = e^{\int_0^t \nabla^2 V(\mathbf{x}(\tau)) d\tau} \nabla \Lambda_0(\mathbf{x}_0)
\end{align*}
The optimal pattern evolution is then carried out by the flow 
\begin{align*}
     \frac{d}{dt}\mathbf{x}(t) = - \frac{1}{\gamma_{\ctrl}} e^{\int_0^t \nabla^2 V(\mathbf{x}(\tau)) d\tau} \nabla \Lambda_0(\mathbf{x}_0) - \nabla V(\mathbf{x}(t))
\end{align*}
where as before $\Lambda_0$ is specified by the terminal constraints~$\phi_0$ and~$\phi^*$ on the pattern evolution. 

We now perform a perturbation analysis of the optimal pattern evolution with respect to the nominal case. To this end, we let $\mathbf{x}^{(\varepsilon)}$ be the transport path in the presence of a potential $\varepsilon V$ and $\mathbf{x}^{(0)}$ the nominal transport path. We then obtain
\begin{align*}
    \frac{d}{dt} \left( \mathbf{x}^{(\varepsilon)} - \mathbf{x}^{(0)} \right) = - \frac{1}{\gamma_{\ctrl}} e^{\varepsilon \int_0^t \nabla^2 V(\mathbf{x}^{(\varepsilon)}(\tau)) d\tau} \nabla \Lambda_0^{(\varepsilon)}(\mathbf{x}^{(\varepsilon)}_0) - \varepsilon \nabla V(\mathbf{x}^{(\varepsilon)}(t)) + \frac{1}{\gamma_{\ctrl}} e^{\varepsilon \int_0^t \nabla^2 V(\mathbf{x}(\tau)) d\tau} \nabla \Lambda^{(0)}_0(\mathbf{x}^{(0)}_0)
\end{align*}
From a Taylor expansion and ignoring $\mathcal{O} \left(\varepsilon^2 \right)$ terms above, we obtain
\begin{align*}
    \frac{d}{dt} \left( \frac{\mathbf{x}^{(\varepsilon)}(t) - \mathbf{x}^{(0)}(t)}{\varepsilon} \right) = - \frac{1}{\gamma_{\ctrl}} \Lambda'_0 \left(\mathbf{x}^{(0)}_0 \right) - \frac{1}{\gamma_{\ctrl}} \int_0^t \nabla^2 V (\mathbf{x}^{(0)}(\tau)) d\tau ~ \nabla \Lambda_0 \left(\mathbf{x}^{(0)}_0 \right) - \nabla V \left(\mathbf{x}^{(0)}(t) \right)
\end{align*}
where $\Lambda'_0 = \lim_{\varepsilon \rightarrow 0} \frac{\Lambda^{(\varepsilon)}_0 - \Lambda^{(0)}_0}{\varepsilon}$. We note that the expression on the right hand side above is evaluated along the nominal transport path~$\mathbf{x}^{(0)}$. As seen previously, the nominal pattern evolution occurs along straight line paths and we infer from the above that the curvature of the transport paths is determined by the Hessian of the potential~$U$~\cite{MS-AL-VB:24}.

\begin{figure}[!h]
    \centering
    \includegraphics[width=1.0\linewidth]{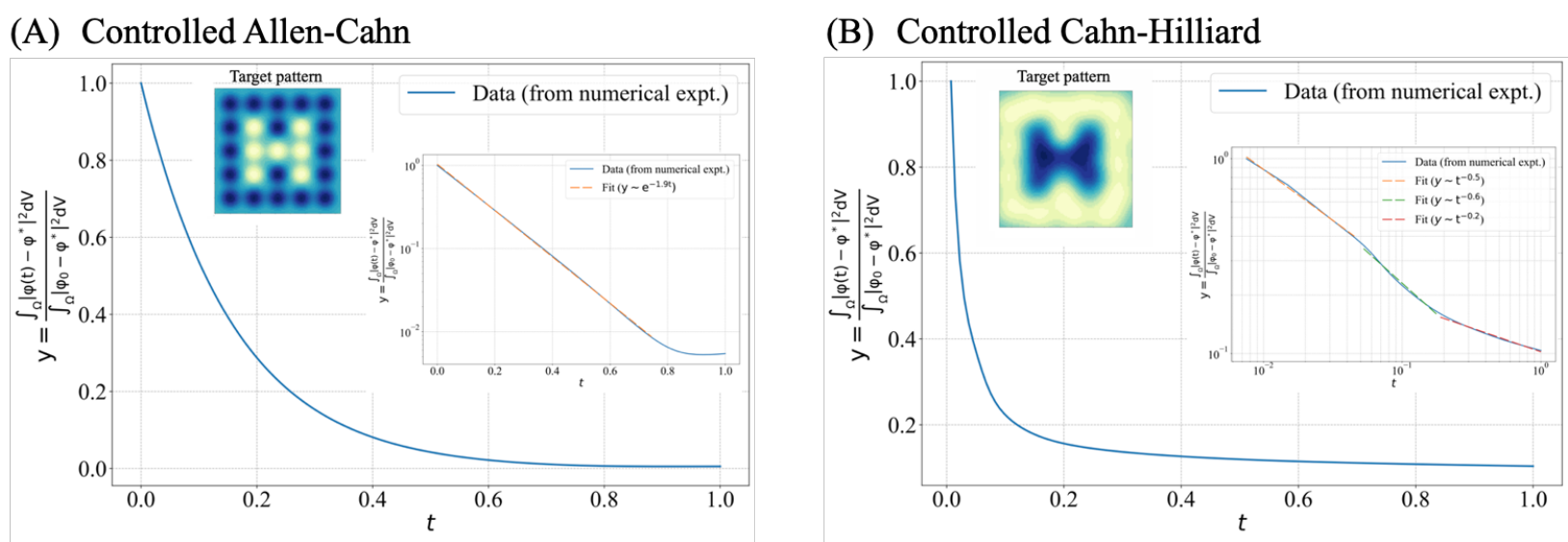}
    \caption{{\bf Targeted phase separation.} {\bf (A)} Model A: Allen-Cahn equation. The Eulerian pattern error and the target pattern `H' (inset) is shown on the rightmost panel (inset showing the error on a logarithmic scale indicating exponential decay). 
    {\bf (B)} Model B: Cahn-Hilliard equation. The Eulerian pattern error and the target pattern `H' (inset) is shown on the rightmost panel (inset showing the error on a log-log scale indicating power law decay).}
    \label{fig:SI_AC_CH}
\end{figure}

\section{Numerical formalism}
We first provide the numerical formalism showing the equivalence between a lagrangian scheme and the eulerian scheme. Next, utilizing the concept of gradient flows, we show how to simulate eulerian pdes via lagrangian coordinates. Finally, we perform numerical experiments where we consider the following examples: (a) Allen-Cahn equation describing the dynamics of unconserved phase separating systems. (b) Cahn-Hilliard equation describing the dynamics of conserved phase separating system (c)Conserved eulerian PDE describing the dynamics of thin films. (d) Reaction-diffusion systems. (e) Phenomenological system for cell-fate decision making.
\subsubsection*{Lagrangian description and optimal transport}
Consider a system of~$N$ active particles with positional configuration $\xvec(t)= ( \xvec_1(t), \xvec_2(t),..., \xvec_N(t) )$, with $\xvec_i(t) \in \real^{d_x}$, and state $\phivec(t) = ( \phi_1(t), \ldots, \phi_N(t) )$ with $\phi_i \in \real^{d_{\phi}}$ at time $t \in [0, T]$ for $i \in \lbrace 1, \ldots, N \rbrace$. Inspired by theories of non-equilibrium statistical physics \cite{RZ:01}, we assume their dynamics is described by system of overdamped Langevin equations
\begin{align} \label{eq:N-body_system}
\begin{aligned}
     \dot{\xvec}(t) &= \muvec(t, \xvec, \phivec) + \ctrlvec(t) + \sqrt{2D_\xi} \xivec(t), \quad  \xvec(0) = \xvec_0, \\
     \dot{\phivec}(t) &= \nuvec(t, \xvec, \phivec) + \vvec(t) + \sqrt{2D_\eta} \etavec(t), \quad \phivec(0) = \phivec_0,
\end{aligned}
\end{align}
where $\muvec(t, \xvec, \phivec)$ and $\nuvec(t, \xvec, \phivec)$ are the known drift vector fields corresponding to $\xvec(t)$ and $\phivec(t)$ respectively. The diffusion terms are given by $\sqrt{2D_\xi} \xivec(t)$ and $\sqrt{2D_\eta} \etavec(t)$, where $D_\xi$ and $D_\eta$ dictate the strength of fluctuations and $ \xivec(t)$ and $\etavec(t)$ correspond to fluctuations with gaussian-white noise statistics. We also augment the drift and diffusion terms with controls $\ctrlvec(t)$ and $\vvec(t)$ which enable the steering of the configuration-state pair~$(\xvec, \phivec)$. The corresponding Fokker-Planck (FP) equation is given by 
\begin{align} \label{eq:FP}
\begin{aligned}
    \frac{\partial p}{\partial t} = D_\xi \sum_{i=1}^N \frac{\partial^2 p}{\partial \vecx_i^2} + D_\eta \sum_{i=1}^N \frac{\partial^2 p}{\partial \phi_i^2} - \sum_{i=1}^N \frac{\partial}{\partial \x_i} \cdot( p (\mu_i + \ctrl_i) ) - \sum_{i=1}^N \frac{\partial}{\partial \phi_i} \cdot (p (\nu_i + \vctrl_i)) \\ 
    p(0, \xvec, \phivec) = p_0(\xvec, \phivec)
\end{aligned}
\end{align}
where $p(t,\xvec,\phivec)$ is the probability density that the configuration is located at $(\xvec, \phivec)$, with the property $\int_{\real^{N d_x}\times \real^{N d_{\phi}}} p(t,\xvec,\phivec) d\xvec d\phivec =1$, $\forall t\in [0,T]$. The goal of a generative model is to sample from a target probability distribution~$p^*$ on the Lagrangian coordinates $(\xvec,\phivec)$. To this end, we formulate the generative model via a minimum-work optimal transport problem of steering the initial probability distribution $p_0$ towards the target probability distribution $p^*$ over a time horizon $[0,T]$, given by
\begin{align*}
    \min_{\ctrlvec, \vvec} \int_0^T \int p(t,\xvec,\phivec) \left( \frac{\gamma_u}{2} \left\| \ctrlvec(t) \right\|^2 + \frac{\gamma_v}{2} \left\| \vvec(t) \right\|^2 \right) d\xvec d\phivec + D(p_T, p^*) 
    \quad \text{s.t.} ~ \text{Equation}~\eqref{eq:FP}~\text{holds}
\end{align*}
where the terminal penalty~$D(p_T, p^*)$ on $p_T(\xvec,\phivec) =  p(T,\xvec,\phivec)$ penalizes its distance from $p^*$. The action functional for the above problem is given by
\begin{align*}
    \mathcal{S} = &\int_0^T \int p(t,\xvec,\phivec) \left( \frac{\gamma_u}{2} \left\| \ctrlvec(t) \right\|^2 + \frac{\gamma_v}{2} \left\| \vvec(t) \right\|^2 \right) d\xvec d\phivec + D(p_T, p^*) \\ 
    &+ \int_0^T \int \mathcal{F} \left( \frac{\partial p}{\partial t} - D_\xi \sum_{i=1}^N \frac{\partial^2 p}{\partial \vecx_i^2} - D_\eta \sum_{i=1}^N \frac{\partial^2 p}{\partial \phi_i^2} + \sum_{i=1}^N \frac{\partial}{\partial \x_i} \cdot( p (\mu_i + \ctrl_i) ) + \sum_{i=1}^N \frac{\partial}{\partial \phi_i} \cdot (p (\nu_i + \vctrl_i))  \right) d\xvec d\phivec 
\end{align*}
The optimal controls~$\ctrlvec^*, \vvec^*$ are obtained by minimizing the above functional w.r.t.~$\ctrlvec, \vvec$ and satisfy $\ctrl^*_i = - \frac{1}{\gamma_u} \frac{\partial \mathcal{F}^*}{\partial \x_i}$ and $\mathrm{v}^*_i = - \frac{1}{\gamma_v} \frac{\partial \mathcal{F}^*}{\partial \phi_i}$, where $\mathcal{F}^*$ is obtained by setting the first variation of $S$ w.r.t.~$p$ to zero and satisfies
\begin{align}
\begin{aligned}
    \frac{\partial \mathcal{F}}{\partial t} + D_\xi \sum_{i=1}^N \frac{\partial^2 \mathcal{F}}{\partial \vecx_i^2} + D_\eta \sum_{i=1}^N \frac{\partial^2 \mathcal{F}}{\partial \phi_i^2} - \frac{1}{2\gamma_u} \sum_{i=1}^N \left| \frac{\partial \mathcal{F}}{\partial \vecx_i} \right|^2 - \frac{1}{2\gamma_v} \sum_{i=1}^N \left| \frac{\partial \mathcal{F}}{\partial \phi_i} \right|^2 + \sum_{i=1}^N \frac{\partial \mathcal{F}}{\partial \vecx_i} \cdot \mu_i + \sum_{i=1}^N \frac{\partial \mathcal{F}}{\partial \phi_i} \cdot \nu_i = 0 \\
    \mathcal{F}(T, \xvec, \phivec) = - \frac{\delta D}{\delta p}(p_T, p^*)
\end{aligned}
\end{align}
which is the Hamilton-Jacobi-Bellman equation corresponding to the minimum-work optimal transport of the probability distribution~$p(t,\xvec,\phivec)$ from $p_0(\xvec,\phivec)$ at time $t=0$ with a terminal penalty~$D(p_T, p^*)$ on $p_T(\xvec,\phivec) =  p(T,\xvec,\phivec)$ penalizing its distance from $p^*$.

\subsubsection*{Eulerian description of pattern evolution}
We note that Equation~\eqref{eq:N-body_system} describes the evolution of the dynamical system in Lagrangian coordinates but a pattern is more appropriately described in an Eulerian sense. To this end, we define a coarse-grained Eulerian description~$\varphi$ of the state~$(\xvec, \phivec)$ via a kernel $K$ as follows
\begin{align*}
    \phi(t,s,\zvec) &= \sum_{i=1}^N K(s, \zvec - \xvec_i(t)) \phi_i(t)
\end{align*}
where $s$ is the scale parameter and $z \in \real^{d_x}$ is the Eulerian coordinate. The sample-averaged pattern is then given by
\begin{align*}
    \bar{\phi}(t,s,z) &= \mathbb{E} \left[ \sum_{i=1}^N K(s, z - \x_i(t)) \phi_i(t) \right] \\
    &= \int p(t, x_1, \ldots, x_N, \phi_1, \ldots, \phi_N) \left(  \sum_{i=1}^N K(s, z - x_i) \phi_i \right) dx_1 \ldots dx_N d\phi_1 \ldots d\phi_N,
\end{align*}
which evolves in time as follows
\begin{align*}
    \frac{\partial \bar{\phi}}{\partial t}(t,s,z) &= \int \frac{\partial p}{\partial t}(t, x_1, \ldots, x_N, \phi_1, \ldots, \phi_N) \left(  \sum_{i=1}^N K(s, z - x_i) \phi_i \right) dx_1 \ldots dx_N d\phi_1 \ldots d\phi_N \\
    &= \int p(t, x_1, \ldots, x_N, \phi_1, \ldots, \phi_N) \left( \sum_{i=1}^N D_\xi \Delta K(s, z - x_i) \phi_i \right) dx_1 \ldots dx_N d\phi_1 \ldots d\phi_N \\
    &\quad - \int p(t, x_1, \ldots, x_N, \phi_1, \ldots, \phi_N) \left( \sum_{i=1}^N (\mu_i + \ctrl_i) \cdot \nabla K(s, z - x_i) \phi_i \right) dx_1 \ldots dx_N d\phi_1 \ldots d\phi_N \\
    &\quad + \int p(t, x_1, \ldots, x_N, \phi_1, \ldots, \phi_N) \left( \sum_{i=1}^N (\nu_i + \vctrl_i) K(s, z - x_i) \right) dx_1 \ldots dx_N d\phi_1 \ldots d\phi_N \\
    &= D_\xi \Delta \bar{\phi}(t,z) \\
    &\quad - \nabla \cdot \left( \int p(t, x_1, \ldots, x_N, \phi_1, \ldots, \phi_N) \left( \sum_{i=1}^N (\mu_i + \ctrl_i) K(s, z - x_i) \phi_i \right) dx_1 \ldots dx_N d\phi_1 \ldots d\phi_N \right) \\
    &\quad + \int p(t, x_1, \ldots, x_N, \phi_1, \ldots, \phi_N) \left( \sum_{i=1}^N (\nu_i + \vctrl_i) K(s, z - x_i) \right) dx_1 \ldots dx_N d\phi_1 \ldots d\phi_N \\
    &=  D_\xi \Delta \bar{\phi}(t,z) + \nabla \cdot (J + J_{\ctrl}) + R + S
\end{align*}

where the flux fields~$J, J_{\ctrl}$ and reaction fields~$R, S$ are given by
\begin{align} \label{eq:eulerian_ctrl}
\begin{aligned}
    J(t,s,z) &= - \int p(t, x_1, \ldots, x_N, \phi_1, \ldots, \phi_N) \left( \sum_{i=1}^N \mu_i K(s, z - x_i) \phi_i \right) dx_1 \ldots dx_N d\phi_1 \ldots d\phi_N \\
    J_{\ctrl}(t,s,z) &= - \int p(t, x_1, \ldots, x_N, \phi_1, \ldots, \phi_N) \left( \sum_{i=1}^N \ctrl_i K(s, z - x_i) \phi_i \right) dx_1 \ldots dx_N d\phi_1 \ldots d\phi_N \\
    R(t,s,z) &= \int p(t, x_1, \ldots, x_N, \phi_1, \ldots, \phi_N) \left( \sum_{i=1}^N \nu_i K(s, z - x_i) \right) dx_1 \ldots dx_N d\phi_1 \ldots d\phi_N \\
    S(t,s,z) &= \int p(t, x_1, \ldots, x_N, \phi_1, \ldots, \phi_N) \left( \sum_{i=1}^N \vctrl_i K(s, z - x_i) \right) dx_1 \ldots dx_N d\phi_1 \ldots d\phi_N
\end{aligned}
\end{align}
\subsubsection*{Converting Eulerian PDEs into Lagrangian form for gradient flows}
In the deterministic limit, the flux and reaction terms are given by
\begin{align*}
    J(t,s,z) &= - \sum_{i=1}^N \mu_i K(s, z - \x_i) \phi_i \\
    R(t,s,z) &= \sum_{i=1}^N \nu_i K(s, z- \x_i) 
\end{align*}
The time derivative of the free energy functional~$E$ is then given by
\begin{align*}
    \frac{dE}{dt} &= \int_{\Omega} \frac{\delta E}{\delta \varphi} \frac{\partial \phi}{\partial t} d^2 \mathbf{z}
                = \int_{\Omega} \frac{\delta E}{\delta \varphi} \nabla \cdot J d^2 \mathbf{z} + \int_{\Omega} \frac{\delta E}{\delta \varphi} R d^2 \mathbf{z} \\
                &= - \int_{\Omega} \nabla \left( \frac{\delta E}{\delta \varphi} \right) \cdot J d^2 \mathbf{z} + \int_{\Omega} \frac{\delta E}{\delta \varphi} R d^2 \mathbf{z} \\
                &= \sum_{i=1}^N \int_{\Omega} \phi_i K(s, z - \x_i) \nabla \left( \frac{\delta E}{\delta \varphi} \right) \cdot  \mu_i  d^2 \mathbf{z} + \sum_{i=1}^N \int_{\Omega}  K(s, z- \x_i) \frac{\delta E}{\delta \varphi} \nu_i d^2 \mathbf{z} \\
                &= \sum_{i=1}^N \underbrace{\int_{\Omega} \phi_i K(s, z - \x_i) \nabla \left( \frac{\delta E}{\delta \varphi}\right) d^2 \mathbf{z}}_{=\frac{\partial E}{\partial \x_i}}   \cdot  \mu_i + \sum_{i=1}^N \underbrace{ \int_{\Omega}  K(s, z- \x_i) \frac{\delta E}{\delta \varphi} d^2 \mathbf{z} }_{= \frac{\partial E}{\partial \phi_i}} \nu_i \\
                &= \sum_{i=1}^N \frac{\partial E}{\partial \x_i} \cdot  \mu_i + \sum_{i=1}^N \frac{\partial E}{\partial \phi_i} \nu_i
\end{align*}
We obtain a gradient flow on~$E$ by setting
\begin{align*}
    \mu_i &= - \frac{\partial E}{\partial \x_i} = - \int_{\Omega} \phi_i K(s, z - \x_i) \nabla \left( \frac{\delta E}{\delta \varphi}\right) d^2 \mathbf{z} \\
    \nu_i &= - \frac{\partial E}{\partial \phi_i} = - \int_{\Omega}  K(s, z- \x_i) \frac{\delta E}{\delta \varphi} d^2 \mathbf{z} 
\end{align*}
\subsubsection*{Optimal control algorithm for Lagrangian dofs}
\begin{figure}
\centering
\includegraphics[width=0.40\linewidth]{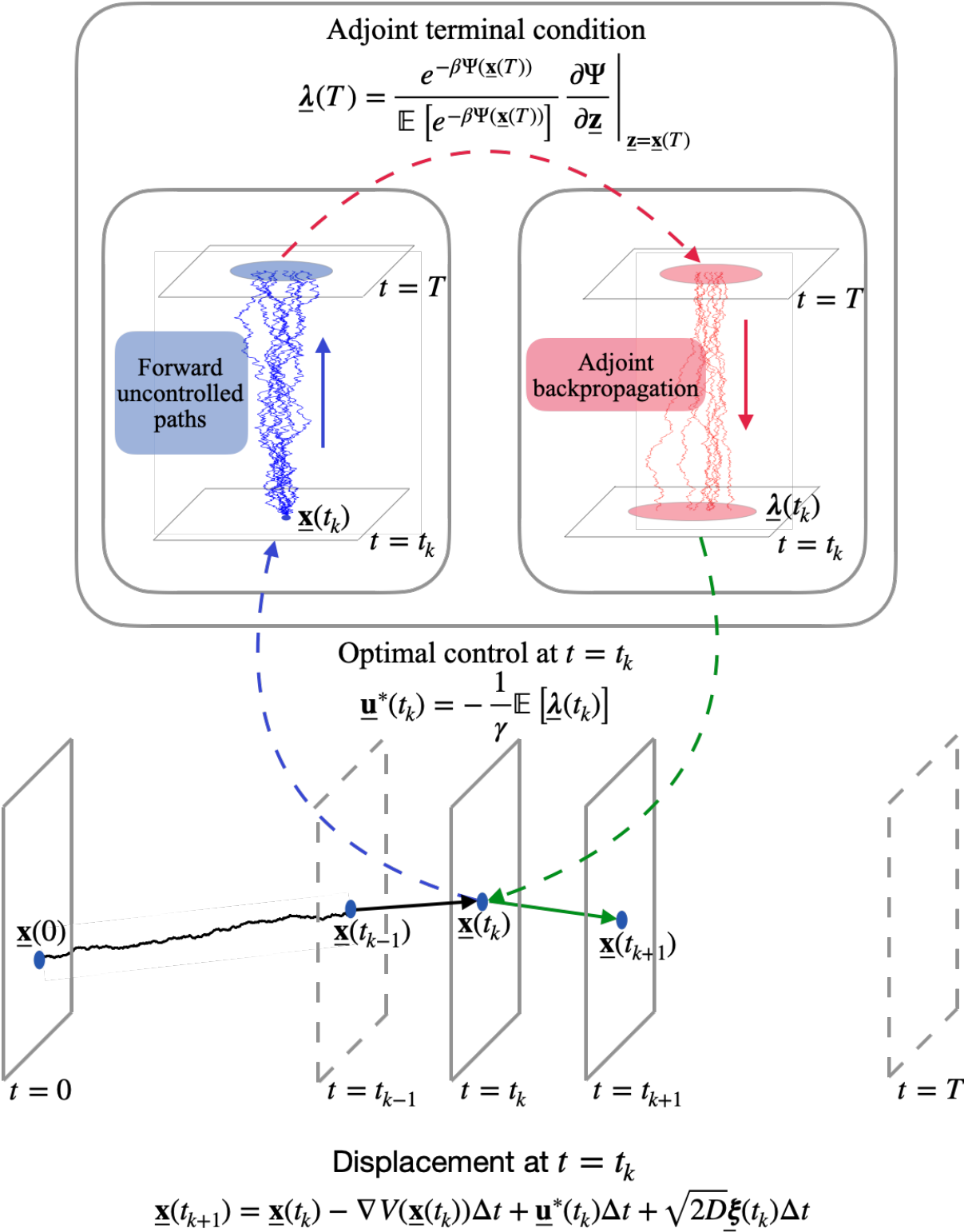} \hspace{0.03\linewidth}
\begin{minipage}{0.46\linewidth}
\vspace{-3.3in}
\begin{algorithm}[H] 
\caption{{\bf Adjoint-based Path Integral Control}}
\textbf{Input:} Current time~$t$, state $\xvec(t)$, number of sampling paths~$n$ \\
\textbf{For every time $t$:}
\begin{algorithmic}[1]
	\State Obtain $n$ independent Brownian noise sequences for the interval $t_0$ to $T$
	\State Integrate SDE  with generated Brownian noise sequences to obtain $n$ uncontrolled paths with initial condition $\xvec(t)$
	\State Integrate adjoint ODE with terminal condition for $\lambdavec(T)$ obtained from the terminal state $\xvec(T)$ of each uncontrolled path to obtain $n$ samples of~$\lambdavec(t)$
        \State Obtain optimal control $\ctrlvec^*(t)$ by sample averaging
\end{algorithmic}
	\label{alg:}
\end{algorithm}
\end{minipage}
\caption{{\bf The adjoint method for stochastic optimal control based on the Feynman-Kac path integral formalism.}
{\bf (Left panel)} Schematic illustrating the adjoint method for finite time horizon stochastic optimal control. A controlled trajectory is shown on the left, with time on the vertical axis. The inlay illustrates the computational steps
performed at time $t_0$, $0\leq t_0 \leq T$, to obtain the instantaneous optimal control input $\ctrlvec^*(t_0)$. The adjoint method based on the Feynman-Kac path integral involves forward propagating several uncontrolled paths 
(solutions to the SDE $\dot{\xvec}(t) = -\nabla V(\xvec(t)) + \sqrt{2D} \xivec(t)$)
from time $t_0$ to~$T$ originating from the current state $\xvec(t_0)$ with independently sampled Brownian noise sequences. The resulting terminal states $\xvec(T)$, set the terminal conditions on the (adjoint) co-state $\lambdavec(T)$,
as the Boltzmann-weighted gradient of the terminal cost $\Psi$ evaluated at $\xvec(T)$, which are then backpropagated through the adjoint ODE $\dot{\lambdavec}(t) = - \nabla^2 V (\xvec(t)) \lambdavec(t)$ from time $T$ to $t_0$, resulting in one backpropagated adjoint path for each forward uncontrolled path. The optimal control $\ctrlvec^*(t_0)$ is then obtained from the sample average of the backpropagated adjoint co-states $\lambdavec(t_0)$.
The control of interacting active particle systems involves the integration of many-body dynamics with optimal control, i.e., the evaluation of many-body interaction potentials and its derivatives, integration of high-dimensional ODEs and SDEs and parallelizable implementation of the (stochastic) adjoint method based on the Feynman-Kac path integral formalism.
In our implementation, the (first and higher order) derivatives of the interaction potential are computed by automatic differentiation within the framework supplied by JAX~\cite{jax2018github}. High-dimensional ODE and SDE integrations are performed in Diffrax~\cite{kidger2021on}, a JAX-based library for numerical integration of differential equations. 
For the implementation of the sampling-based Feynman-Kac path integral adjoint algorithm, we utilize the automatic vectorization functionality provided by ${\rm vmap}$ in JAX in combination with ODE/SDE integration in Diffrax.
{\bf (Right panel)} Table containing the Adjoint-based Path Integral Control algorithm.} 
\label{fig:adjoint}
\end{figure}

In all the numerical experiments, the infinite dimensional forward eulerian PDEs were converted to finite dimensional lagrangian dofs moving on a complex landscape (see figure 3 in main text). We then utilize our previous work \cite{SS-VK-LM:23}, where we developed the custom APIC algorithm based stochastic optimal control to navigate lagrangian dofs optimally. The practical implementation is carried out utilizing the automatic differentiation framework provided by JAX and its subsidiaries, as shown in~Figure~\ref{fig:adjoint} (adapted from~\cite{SS-VK-LM:23}).

\section{Targeted assembly of droplets}
In reference to the class of models considered in the main text, i.e., Model A and Model B, we now consider Model B' which corresponds to the dewetting of a fluid on a substrate where the mobility is given by $m(\varphi)=\varphi^3$, and the dewetting velocity field is given by $\mathbf{v}=-\nabla \left(\frac{\delta E }{\delta \varphi} \right)$, where $E(\varphi)=\int_{\Omega} \left[ U(\varphi(\mathbf{z})) + \frac{\epsilon}{2} \left\| \nabla \varphi(\mathbf{z}) \right\|^2 \right] d^2 \mathbf{z}$ and $\phi(t,\mathbf{z})$ is the height of a thin film of liquid at a point $\mathbf{z}$ at time $t$. 

\subsection*{Derivation of the thin film equation}
Here derive the thin film droplet equation for completion. Consider a thin liquid film of thickness \(\phi(x, y, t)\) spreading over a solid substrate. The film occupies the region \(0 \leq z \leq \phi(x, y, t)\), where \(z\) is the coordinate normal to the substrate. The \(x\) and \(y\) coordinates lie along the substrate. \\

\noindent We make the following assumptions regarding the fluid and flow properties:
\begin{enumerate}
    \item The fluid is incompressible and Newtonian with density \(\rho\) and dynamic viscosity \(\mu\).
    \item The flow is laminar and characterized by slow velocities (low Reynolds number), allowing us to neglect inertial terms.
    \item The film is thin: \(\phi \ll L\), where \(L\) is a characteristic length in the \(x\) and \(y\) directions.
    \item Pressure variations in the \(z\)-direction are hydrostatic.
    \item Surface tension \(\gamma\) contributes to the pressure via curvature effects.
\end{enumerate}

\noindent For an incompressible Newtonian fluid, the Navier-Stokes equations are:
\begin{align}
    &\nabla \cdot \bm{v} = 0, \label{continuity} \\
    &\rho \left( \frac{\partial \bm{v}}{\partial t} + \bm{v} \cdot \nabla \bm{v} \right) = -\nabla p + \mu \nabla^2 \bm{v} + \rho \bm{g}, \label{momentum}
\end{align}
where \(\bm{v} = (u, v, w)\) is the velocity vector, \(p\) is the pressure, and \(\bm{g}\) is the gravitational acceleration vector. Under the lubrication approximation for thin films, (a) The flow is predominantly in the \(x\) and \(y\) directions: \(u\) and \(v\) are much larger than \(w\), (b) Gradients in the \(z\)-direction dominate: \(\frac{\partial}{\partial z} \gg \frac{\partial}{\partial x}, \frac{\partial}{\partial y}\), (c) Inertial terms are negligible compared to viscous terms. Hence, the \(z\)-component of the momentum equation simplifies to:
\begin{equation}
    0 = -\frac{\partial p}{\partial z} - \rho g, \label{z-momentum}
\end{equation}
where \(g\) is the acceleration due to gravity in the \(-z\) direction. Integrating Equation \eqref{z-momentum} with respect to \(z\):
\begin{equation}
    p(x, y, z, t) = p_s(x, y, t) - \rho g (\phi - z), \label{pressure}
\end{equation}
where \(p_s(x, y, t)\) is the pressure at the free surface \(z = \phi\). The \(x\)-component of the momentum equation simplifies to:
\begin{equation}
    0 = -\frac{\partial p}{\partial x} + \mu \frac{\partial^2 u}{\partial z^2}. \label{x-momentum}
\end{equation}
Similarly for the \(y\)-component:
\begin{equation}
    0 = -\frac{\partial p}{\partial y} + \mu \frac{\partial^2 v}{\partial z^2}. \label{y-momentum}
\end{equation}
From~\eqref{pressure}, we have:
\begin{equation}
    \frac{\partial p}{\partial x} = \frac{\partial p_s}{\partial x} + \rho g \frac{\partial \phi}{\partial x}, \quad \frac{\partial p}{\partial y} = \frac{\partial p_s}{\partial y} + \rho g \frac{\partial \phi}{\partial y}. \label{pressure-gradient}
\end{equation}
We impose a no-slip boundary condition at $z=0$, i.e., \(u(x,y,0,t) 0,~v(x,y,0,t) = 0\) along with a stress-free condition at the free surface ($z=\phi$) \(\frac{\partial u}{\partial z}(x,y,\phi,t) = \frac{\partial v}{\partial z}(x,y,\phi,t) = 0\). We now solve for the velocity profiles. Substituting~\eqref{pressure-gradient} into~\eqref{x-momentum}:
\begin{equation}
    \mu \frac{\partial^2 u}{\partial z^2} = -\left( \frac{\partial p_s}{\partial x} + \rho g \frac{\partial \phi}{\partial x} \right). \label{x-momentum-simplified}
\end{equation}
Integrating once with respect to \(z\):
\begin{equation}
    \mu \frac{\partial u}{\partial z} = -\left( \frac{\partial p_s}{\partial x} + \rho g \frac{\partial \phi}{\partial x} \right) z + C_1. \label{first-integration}
\end{equation}
Applying the boundary condition at \(z = \phi\):
\begin{equation}
    \left. \frac{\partial u}{\partial z} \right|_{z = \phi} = 0 \implies C_1 = \left( \frac{\partial p_s}{\partial x} + \rho g \frac{\partial \phi}{\partial x} \right) \phi. \label{constant-C1}
\end{equation}
Integrating again with respect to \(z\):
\begin{equation}
    \mu u = -\frac{1}{2} \left( \frac{\partial p_s}{\partial x} + \rho g \frac{\partial \phi}{\partial x} \right) z^2 + C_1 z + C_2. \label{second-integration}
\end{equation}
Applying the boundary condition at \(z = 0\):
\begin{equation}
    u(z=0) = 0 \implies C_2 = 0. \label{constant-C2}
\end{equation}
Substituting \(C_1\) and \(C_2\), the velocity profile in the \(x\)-direction is:
\begin{equation}
    u(z) = \frac{1}{\mu} \left( \frac{\partial p_s}{\partial x} + \rho g \frac{\partial \phi}{\partial x} \right) \left( \phi z - \frac{1}{2} z^2 \right). \label{velocity-profile}
\end{equation}
Similarly, the velocity profile in the \(y\)-direction is:
\begin{equation}
    v(z) = \frac{1}{\mu} \left( \frac{\partial p_s}{\partial y} + \rho g \frac{\partial \phi}{\partial y} \right) \left( \phi z - \frac{1}{2} z^2 \right). \label{velocity-profile-y}
\end{equation}
The continuity equation for incompressible flow is:
\begin{equation}
    \frac{\partial u}{\partial x} + \frac{\partial v}{\partial y} + \frac{\partial w}{\partial z} = 0. \label{continuity-expanded}
\end{equation}
Integrating~\eqref{continuity-expanded} over \(z\) from \(0\) to \(\phi\):
\begin{equation}
    \frac{\partial}{\partial t} \int_0^\phi dz + \frac{\partial}{\partial x} \int_0^\phi u \, dz + \frac{\partial}{\partial y} \int_0^\phi v \, dz + \left. w \right|_{z=\phi} - \left. w \right|_{z=0} = 0. \label{integrated-continuity}
\end{equation}
At the substrate (\(z=0\)), there is no penetration:
\begin{equation}
    w(z=0) = 0.
\end{equation}
At the free surface (\(z=\phi\)), the kinematic boundary condition relates the vertical velocity to the rate of change of \(\phi\):
\begin{equation}
    w(z=\phi) = \frac{\partial \phi}{\partial t} + u(z=\phi) \frac{\partial \phi}{\partial x} + v(z=\phi) \frac{\partial \phi}{\partial y}.
\end{equation}
Assuming \(u(z=\phi)\) and \(v(z=\phi)\) are negligible (due to the stress-free condition and thin film assumption), we simplify:
\begin{equation}
    w(z=\phi) = \frac{\partial \phi}{\partial t}.
\end{equation}
Substitute back into Equation \eqref{integrated-continuity}:
\begin{equation}
    \frac{\partial \phi}{\partial t} + \frac{\partial}{\partial x} \int_0^\phi u \, dz + \frac{\partial}{\partial y} \int_0^\phi v \, dz = 0. \label{mass-conservation}
\end{equation}
We now obtain the volumetric flow rates. Define the volumetric flow rates per unit width:
\begin{equation}
    q_x = \int_0^\phi u \, dz, \quad q_y = \int_0^\phi v \, dz. \label{flow-rates}
\end{equation}
Compute \(q_x\) using Equation \eqref{velocity-profile}:
\begin{align}
    q_x &= \int_0^\phi \frac{1}{\mu} \left( \frac{\partial p_s}{\partial x} + \rho g \frac{\partial \phi}{\partial x} \right) \left( \phi z - \frac{1}{2} z^2 \right) dz \\
    &= \frac{1}{\mu} \left( \frac{\partial p_s}{\partial x} + \rho g \frac{\partial \phi}{\partial x} \right) \left[ \frac{\phi z^2}{2} - \frac{z^3}{6} \right]_0^\phi \\
    &= \frac{\phi^3}{3 \mu} \left( \frac{\partial p_s}{\partial x} + \rho g \frac{\partial \phi}{\partial x} \right). \label{q_x}
\end{align}
Similarly for \(q_y\):
\begin{equation}
    q_y = \frac{\phi^3}{3 \mu} \left( \frac{\partial p_s}{\partial y} + \rho g \frac{\partial \phi}{\partial y} \right). \label{q_y}
\end{equation}
For small slopes (\(|\nabla \phi| \ll 1\)), the curvature of the free surface is approximated by:
\begin{equation}
    \kappa \approx -\nabla^2 \phi. \label{curvature}
\end{equation}
The Laplace pressure at the free surface is then:
\begin{equation}
    p_s = -\gamma \nabla^2 \phi. \label{surface-pressure}
\end{equation}
Substitute Equations \eqref{q_x}, \eqref{q_y}, and \eqref{surface-pressure} into Equation \eqref{mass-conservation}:
\begin{equation}
    \frac{\partial \phi}{\partial t} + \frac{\partial}{\partial x} \left( \frac{\phi^3}{3 \mu} \left( -\gamma \frac{\partial}{\partial x} \nabla^2 \phi + \rho g \frac{\partial \phi}{\partial x} \right) \right) + \frac{\partial}{\partial y} \left( \frac{\phi^3}{3 \mu} \left( -\gamma \frac{\partial}{\partial y} \nabla^2 \phi + \rho g \frac{\partial \phi}{\partial y} \right) \right) = 0. \label{thin-film-expanded}
\end{equation}
On combining terms we get:
\begin{equation}
    \frac{\partial \phi}{\partial t} = \nabla \cdot \left( \frac{\gamma \phi^3}{3 \mu} \nabla \nabla^2 \phi - \frac{\rho g \phi^3}{3 \mu} \nabla \phi \right), \label{thin-film-equation}
\end{equation}
which variationally can be written as 
\begin{equation}
    \frac{\partial \phi}{\partial t} = -\nabla \cdot \left( \frac{\phi^3}{3 \mu} \nabla \left( \frac{\delta \mathcal{E}}{\delta \varphi} \right) \right). \label{thin-film-variational}
\end{equation}
where the energy functional \(\mathcal{E}[\varphi]\):
\begin{equation}
    \mathcal{E}[\varphi] = \int \left( \frac{\gamma}{2} |\nabla \varphi|^2 + \frac{\rho g}{2} \varphi^2 \right) dx \, dy. \label{energy-functional}
\end{equation}
We can now generalize the energy functional beyond the quadratic nonlinearity in the above case, and consider the energy functionals of the form $\mathcal{E}[\varphi] = \int \left( \frac{\gamma}{2} |\nabla \varphi|^2 + U(\varphi({\bf z}))\right) d^2{\bf z}$, where the $U(\varphi({\bf z}))$ incorporates the attraction and repulsion part of the potential energy density.

\subsection*{Control of thin liquid film dynamics}
\begin{figure}[!h]
    \centering
    \includegraphics[width=1.0\linewidth]{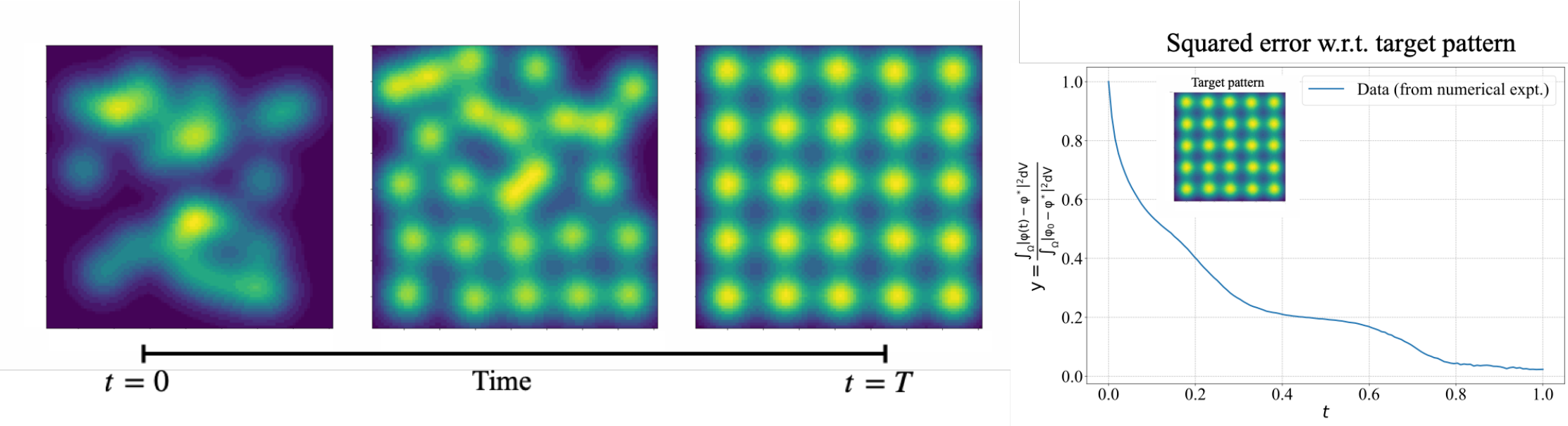}
    \caption{{\bf Targeted assembly of droplets.} Controlling a system of interacting droplets governed by the thin film equation from a random initial configuration. The uncontrolled thin film equation is given by $\frac{\partial \phi}{\partial t} + \nabla \cdot \left( m(\phi)  \mathbf{v} \right)=0$, where $\phi(t, {\bf z})$ is the height of the droplet at time~$t$ and~${\bf z} \in \Omega$, $m(\varphi)=\varphi^3$ is the mobility, and $\mathbf{v}$ is the known vector field given by $\mathbf{v}=-\nabla \frac{\delta E}{\delta \varphi}$, where $E(\varphi)=\int_{\Omega} \left[ U(\varphi(\mathbf{z})) + \frac{\epsilon}{2} \left\| \nabla \varphi(\mathbf{z}) \right\|^2 \right] d^2 \mathbf{z}$ and  $\frac{dU}{d\varphi}=\varphi e^{-\varphi^2}(\varphi^2-\eta)$, where $\epsilon=10^{-3}$ and $\eta=10$. The Eulerian pattern error is shown on the rightmost panel (inset showing the target pattern, a $5\times 5$ square lattice). The error shows resemblance to the decay of the self-intermediate scattering function in super-cooled liquids. }
    \label{fig:droplets}
\end{figure}

The initial distribution is a random distribution given by $ \phi(0,\mathbf{z})$ (see left most panel Figure \ref{fig:droplets}) and the target distribution, $\phi^*$, is a square chequered board pattern. Figure \ref{fig:droplets} shows the sequence of controlled trajectory of the thin film (see Movie3) and the squared error which shows a step-like pattern as observed in the decay of self-intermediate scattering function in super-cooled liquids \cite{LB-GB-JPB-LC-WVS:11}.

\section{Control of reaction-diffusion systems: discrete model}
Consider a pattern composed of three morphogen fields represented as~$\phi(t,{\bf z}) = (\phi_1(t, {\bf z}),\phi_2(t, {\bf z}), \phi_3(t, {\bf z}))$, 
where $\phi_i(t,\mathbf{z})$ is the concentration of the $i$-th morphogen at a location $\zvec \in \Omega \subset \real^2$ at time $t\in [0, T]$. The passive forward reaction-diffusion dynamics for the $i$-th morphogen is given by $\frac{\partial \phi_i}{\partial t} = - \left(\frac{\delta E}{\delta \varphi} \big|_{\phi}\right)_i =  D_i \Delta \phi_i + M_{ijk} \phi_j \phi_k$. The Lagrangian discretization of the coupled PDEs is as follows 
\begin{align*}
    \phi(t, {\bf z}) = \sum_{\nu=1}^{N}\kappa({\bf z} - {\bf x}_\nu(t))\Phi_\nu(t),
\end{align*}
where the Lagrangian coordinates are $\xvec = ({\bf x}_1, \ldots, {\bf x}_N)$ and $\Phi(t) = (\Phi_1(t), \ldots, \Phi_N(t))$. The Lagrangian control problem then becomes
\begin{align*}
    \min_{{\bf u}, v } 
    ~ &\frac{\gamma_u}{2} \sum_{\nu=1}^N \int_{0}^{T} ||{\bf u}_\nu||^2 dt 
    + \frac{\gamma_v}{2} \sum_{\nu=1}^N \sum_{i=1}^3 \int_{0}^{T} {({\rm v}_{\nu})_{i}^2(t)} dt + \Psi(\Phi(T))
    \\ & \text{s.t.}~ 
    \begin{cases}
        &\dot{\bf x}_{\nu} = \mathbf{u}_\nu + \sqrt{2D_\xi} \boldsymbol{\xi}_\nu(t) \\
        &(\dot{\Phi}_\nu)_{i} = M_{ijk} (\Phi_\nu)_{j} (\Phi_\nu)_{k} + ({\rm v}_\nu)_i + \sqrt{2D_\eta} (\eta_\nu)_{i}, \quad \text{for}~~i=1,2,3 \\
    \end{cases}
\end{align*}
The mapping between many-body stochastic Langevin dynamics and deterministic field equations is shown in section S4. Briefly, the control $\mathbf{u}_\nu$ on the velocity of the dof's correspond to the flux control and $R_\nu$ correspond to the reaction control. 

\section{Control of cell fate dynamics: discrete model}
%
Consider~$N$ cells on a two-dimensional planar substrate with spatial location ${\bf x}_i(t) \in \real^2$ at time~$t$. In addition to the physical location of the cell, one also tracks the gene expression profile corresponding to a particular gene, which can take on values $\lbrace -1, 0, 1 \rbrace$ wherein the values~$-1, 1$ correspond to differentiated states and~$0$ corresponds to the undifferentiated pluripotent state. We relax the discreteness constraint and denote the state of the~$i$-th cell at time as $-1\leq \phi_i(t)\leq 1$. The optimal control problem of creating a non-equilibrium phase separation can be cast as 
\begin{align*}
    \min_{{\bf u}, \mathbf{v}} 
    ~ &\frac{\gamma_u}{2} \sum_{i=1}^N \int_{0}^{T} ||{\bf u}_i||^2 dt 
    + \frac{\gamma_v}{2} \sum_{i=1}^N \int_{0}^{T} {{\rm v}_i^2(t)} dt
    + \Psi(\xvec(T), \cvec(T))
    \\ & \text{s.t.}~ 
    \begin{cases}
        &\dot{\bf x}_{i} = - \frac{\partial E}{\partial \xvec_i} (\xvec(t), \phivec(t)) + \mathbf{u}_i(t) + \sqrt{2D_\xi} \boldsymbol{\xi}_i(t) \\
        &\dot \phi_i= - \frac{\partial E}{\partial \phi_i}(\xvec(t), \phivec(t)) + {\rm v}_i(t) + \sqrt{2D_\eta} \eta_i(t) \\
    \end{cases}
\end{align*}
where the interaction potential is given by
\begin{align*}
    E(\xvec, \phivec) &= \sum_{i=1}^N \sum_{j=1}^N \left[ - e^{- \frac{\left\| \mathbf{x}_i - \mathbf{x}_j \right\|^2}{2\ell_1^2}} + \eta (\phi_i - \phi_j)^2 e^{- \frac{\left\| \mathbf{x}_i - \mathbf{x}_j \right\|^2}{2\ell_2^2}} \right] 
\end{align*}
with $\ell_1 > \ell_2$ and $\eta$ (unity in numerics) regulates the strength of the repulsion between cells with different fate. Here, $l_1$ and $l_2$ are the scales of attractive and repulsive interactions with $l_1=1.5, l_1/l_2=5$ is the numerical experiments. We note that the inter-cellular interaction consists of a long-range attraction and short-range repulsion between different cell types. The terminal penalty is given by
\begin{align*}
    \Psi(\xvec, \phivec) &= \frac{\delta_{c}}{4} \sum_{i=1}^{N} \prod_{k=1}^{2} \left( \phi_i - C_k \right)^2  + \frac{\delta_\mu}{2} \sum_{k=1}^2 \sum_{i=1}^N  e^{-\frac{(\phi_i - C_k)^2}{2\sigma^2}} \left\| \mathbf{x}_i - \mu^*_k \right\|^2  
\end{align*}
where $C_k = \lbrace -1, 1 \rbrace$ and the morphogen sets $\mu^*_k$. The first term in~$\Psi$ corresponds to speciation and penalizes undifferentiated terminal states, while the second term corresponds to segregation and penalizes deviation from corresponding target locations for individual cell types.
\begin{figure}[!h]
    \centering
    \includegraphics[width=0.7\linewidth]{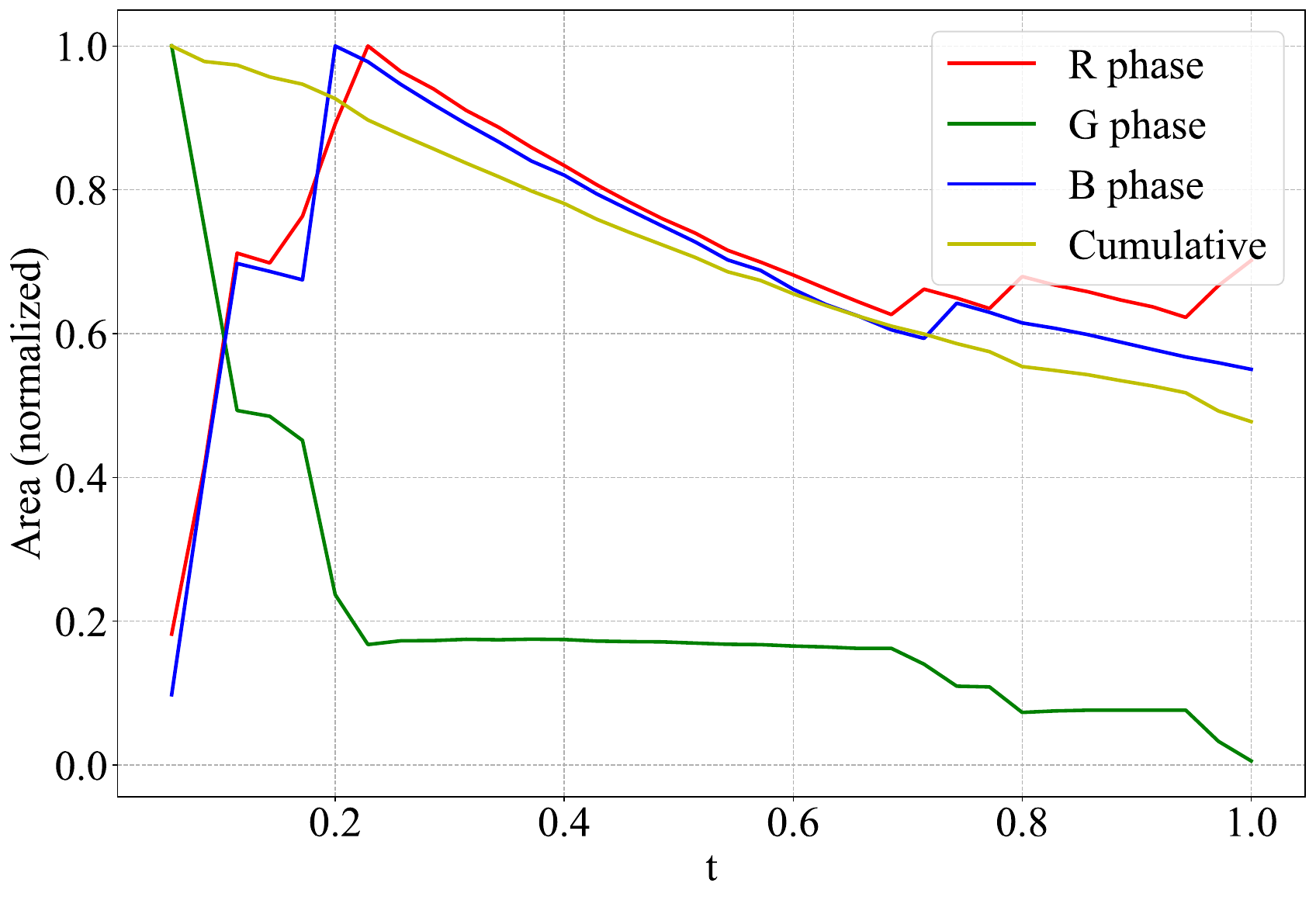}
    \caption{{\bf Quantification of specie area in the cell fate control} The homogeneous state (green) decreases as a function of time, whereas the red and blue differentiated states initially increases and then decreases due to migration of the clusters. The initial increase of the area qualitatively looks similar to the travelling fronts observed during embryonic development \cite{SDT-MV:22}.}
    \label{fig:area_ST}
\end{figure}

\section{Formal derivation of the Hamilton-Jacobi-Bellman equation}
To derive the Hamilton-Jacobi-Bellman (HJB) equation for the given problem, we start with the passive pattern forming dynamics for the nominal case, given by
\begin{align*}
\frac{\partial \phi}{\partial \tau} + \nabla \cdot \left( m(\phi) \mathbf{u} \right) = S
\end{align*}
with initial condition \(\phi(t, \mathbf{z}) = \varphi(\mathbf{z})\). The objective is to minimize the cost functional
\begin{align*}
\mathcal{F}(t, \varphi) = \min_{\mathbf{u}} & \left\{ \frac{\gamma_{\ctrl}}{2} \int_t^T \int_\Omega m(\phi) \|\mathbf{u}\|^2 \, d^2 \mathbf{z} \, d\tau + \frac{\gamma_{\vctrl}}{2} \int_t^T \int_\Omega S^2 \, d^2 \mathbf{z} \, d\tau \right. \nonumber \\
&\left. + \int_\Omega \left[ \Psi(\phi(T, \mathbf{z})) + \frac{\epsilon}{2} \|\nabla \phi(T, \mathbf{z})\|^2 \right] \, d^2 \mathbf{z} \right\}
\end{align*}
where the value function \(\mathcal{F}(t, \varphi)\) describes the minimal cost from time \(t\) to terminal time \(T\). We now consider the change in the value function~$\mathcal{F}$ over an infinitesimal time increment \(d\tau\)
\begin{align} \label{eq:value_func_increment}
\mathcal{F}(t, \varphi) = \min_{\mathbf{u}, S} \left\{ \frac{\gamma_{\ctrl}}{2} \int_\Omega m(\varphi) \|\mathbf{u}\|^2 \, d^2 \mathbf{z} + \frac{\gamma_{\vctrl}}{2} \int_\Omega S^2 \, d^2 \mathbf{z} + \mathcal{F}(t + d\tau, \varphi + d\phi) \right\}
\end{align}
Taylor expanding~$\mathcal{F}(t + d\tau, \varphi + d\phi)$, we get
\begin{align*}
\mathcal{F}(t + d\tau, \varphi + d\phi) \approx \mathcal{F}(t, \varphi) + \frac{\partial \mathcal{F}}{\partial t} d\tau + \int_\Omega \frac{\delta \mathcal{F}}{\delta \varphi} d\phi \, d^2 \mathbf{z}
\end{align*}
Substituting the evolution of the state~$\phi$ (with increment $d\phi = \left( -\nabla \cdot (m(\varphi) \mathbf{u}) + S \right) d\tau$) above, we obtain the following minimization problem for the optimal control vector and reaction fields
\begin{align*}
\min_{\mathbf{u}, S} & \left\{ \frac{\gamma_{\ctrl}}{2} \int_\Omega m(\varphi) \|\mathbf{u}\|^2 \, d^2 \mathbf{z} + \frac{\gamma_{\vctrl}}{2} \int_\Omega S^2 \, d^2 \mathbf{z} \right. \nonumber  \left. + \int_\Omega \nabla \left( \frac{\delta \mathcal{F}}{\delta \varphi} \right) \cdot (m(\varphi) \mathbf{u}) \, d^2 \mathbf{z} + \int_\Omega \frac{\delta \mathcal{F}}{\delta \varphi} S \, d^2 \mathbf{z} \right\}
\end{align*}
The optimal control vector and reaction fields are are then given by
\[
\mathbf{u}^* = -\frac{1}{\gamma_{\ctrl}} \nabla \left( \frac{\delta \mathcal{F}}{\delta \varphi} \right), \quad S^* = - \frac{1}{\gamma_{\vctrl}} \frac{\delta \mathcal{F}}{\delta \varphi}
\]
Substituting the optimal control vector and reaction fields, along with the Taylor expansion, in~\eqref{eq:value_func_increment}, we get
\begin{align*}
\begin{aligned}
    \frac{\partial \mathcal{F}}{\partial t} + \frac{1}{2\gamma_{\ctrl}} \int_\Omega  m(\varphi) \left\| \nabla \left( \frac{\delta \mathcal{F}}{\delta \varphi} \right) \right\|^2 d^2\mathbf{z} + \frac{1}{2 \gamma_{\vctrl}} \int_\Omega \left( \frac{\delta \mathcal{F}}{\delta \varphi} \right)^2 d^2\mathbf{z} = 0& \quad \text{holds~for~any~} t, \varphi \\ 
    \mathcal{F}(T,\varphi) = \int_{\Omega} \left[\Psi(\varphi({\bf z})) + \frac{\epsilon}{2} \left\| \nabla \varphi(\mathbf{z}) \right\|^2 \right] d^2 \mathbf{z}.&
\end{aligned}
\end{align*}

\end{document}